\documentclass[amsmath,amssymb,showpacs,showkeywords]{revtex4}
\usepackage[dvips]{graphicx,color}
\usepackage{times}
\usepackage{mathrsfs}
\usepackage{amsmath}
\usepackage{graphicx}
\usepackage{epsfig}
\usepackage{dcolumn}
\usepackage{bm}
\begin{document}

\title{Azimuthal distribution of Cherenkov photons and corresponding 
electron-positron asymmetry in EASs of different primaries}
\author{G. S. Das\footnote{gsdas@dibru.ac.in}, P. Hazarika\footnote{poppyhazarika1@gmail.com} and U. D. Goswami\footnote{umananda@dibru.ac.in}}
\affiliation{Department of Physics, Dibrugarh University, Dibrugarh 786004, 
Assam, India}

\begin{abstract}
We study the azimuthal distributions of Cherenkov photons in Extensive Air 
Showers (EASs) initiated by $\gamma$-ray, proton and iron primaries of 
different energies incident at various zenith angles over a high altitude
observation level. The azimuthal distributions of electrons and positrons along
with their asymmetric behaviour have also been studied here to understand the 
feature of azimuthal distributions of Cherenkov photons in EASs. The main 
motivation 
behind this study is to see whether the azimuthal distribution of Cherenkov 
photons can provide any means to distinguish the $\gamma$-ray initiated 
showers from that of hadron initiated showers in the ground based $\gamma$-ray 
astronomy experiment. Apart from this, such study is also important to 
understand the natures of $\gamma$-ray and hadronic showers in general. We have 
used the CORSIKA 6.990 simulation package for generating the showers. The 
study shows the double peak nature of the azimuthal distribution of Cherenkov
photons which is due to the separation of electron and positrons in the 
azimuthal plane. The pattern of distribution is more sensitive for the energy
of the primary particle than it's angle of incidence. There is no significant
difference between distributions for $\gamma$-ray and handron initiated 
showers.  
\end{abstract}

\maketitle
\section{Introduction}
\label{sec1}
The earth's atmosphere is opaque to $\gamma$-rays coming from astrophysical
sources. So for the ground based 
detection of VHE $\gamma$-rays from such celestial sources in the energy range 
of few hundred GeV to few TeV, an indirect detection method known as the 
'Atmospheric Cherenkov Technique (ACT)' is used most extensively. This 
technique is based on detection of Cherenkov photons which are emitted in the
atmosphere by the charged particles of Extensive Air Showers (EASs) produced
by the incoming $\gamma$-rays during the interactions with the earth's 
atmospheric particles. However, there is an inherent problem with the ACT that 
the sources which emit $\gamma$-rays also produces Cosmic Rays (CRs). CRs
also produce EASs in the earth's atmosphere and hence the Cherenkov photons 
captured with the ground based detectors contains contributions from both the 
$\gamma$-rays and CRs. Unlike $\gamma$-rays, CRs are charged particles and 
so they are deflected by the intragalactic magnetic fields. As a result of 
which, the CRs when reach the earth, loses their direction of origin. 
Whereas $\gamma$-rays being electrically neutral retains its direction of 
origin. Therefore the detection of $\gamma$-rays can provide the information 
about the direction of their origin. Therefore, for proper estimation of 
different parameters of the incident $\gamma$-rays, the huge background of CRs 
has to be removed from the observed data. Thus the development of effective 
$\gamma$-hadron separation technique is an important issue in the ground based
$\gamma$-ray astronomy experiments, where ACT is being used. In this regard 
the detail study of lateral, temporal and angular distributions of Cherenkov 
photons in EASs of different primaries using Monte Carlo simulation is 
essential for proper disentangle of the $\gamma$-ray showers from the hadronic 
ones. 

In our earlier works \cite{Hazarika, Das} we have already made some detail 
investigations on parameters, viz., density, arrival time and angular 
distributions of Cherenkov photons in EASs of different primaries to 
distinguish $\gamma$-ray initiated showers from that of hadron initiated
showers. Some very interesting works related
to the azimuthal and other angular distributions of Cherenkov photons in EASs 
and corresponding 
electron positron asymmetry has already been carried out using available 
detailed simulation techniques \cite{Homola, Bardan, Cabot, Hillas, Lafebre, 
Nerling}. However, not many studies have been found which focus on for 
distinguishing the 
$\gamma$ and hadron initiated showers, specially over high altitude observation
levels. Hence, in this work we study the azimuthal distribution
of Cherenkov photons in EASs of different primaries incident over a high 
altitude observation level for exploring the possibility of providing 
additional inputs to the effective $\gamma$-hadron separation techniques. 

We have generated EASs for $\gamma$-rays, proton and iron nuclei with 
different energies and angle of inclination with the zenith. The simulation 
has been done by using CORSIKA 6.990 package \cite{Heck}. CORSIKA is a detailed
Monte Carlo simulation code to study the evolution and properties of EASs in 
the atmosphere. Using CORSIKA one can simulate interactions and decays of 
nuclei, hadrons, muons, electrons and photons in the atmosphere up to energies 
of some 10$^{20}$ eV \cite{Knapp}. The availability of seven  high energy 
hadronic interaction models and three low energy hadronic interaction models 
makes the CORSIKA suitable for simulation study of variety of hadronic 
interactions. It uses EGS4 code \cite{Nelson} for the simulation of 
electromagnetic component of the air shower \cite{Knapp}. This paper has been 
organized as follows. The details about the simulation 
process is discussed in the next section. In the section \ref{sec3}, we 
discuss about the analysis of the simulated data and the results. The summary 
and conclusion of the work are made in the section \ref{sec4}.    

\section{Simulation of the extensive air shower}
\label{sec2}
In this simulation work we have used the QGSJETII.3 high energy hadronic 
interaction model and the FLUKA low energy hadronic interaction models out of 
the available options in the CORSIKA 6.990 package. In our earlier works 
\cite{Hazarika, Das} we have already compared the other available high and low 
energy hadronic interaction models available in CORSIKA and found that almost 
all the combinations of high energy and low energy models produces very 
identical results with one another. Thereby revealing almost model independent 
nature of our study, specially for the input parameters used during the 
simulation. We have generated EASs for $\gamma$-ray, proton and iron primaries 
incident vertically as well as inclined at zenith angle 10$^{\circ}$, 
20$^{\circ}$, 30$^{\circ}$ and 40$^{\circ}$. The numbers of showers generated 
at different energies and zenith angles for the three primaries are given in 
Table \ref{tab1}.

\vspace{-0.4cm}
\begin{center}
\begin{table}[ht]
\caption{\label{tab1} Number of showers generated at different energies and at 
different zenith angles for different primary particles.}
\begin{tabular}{ccc}\hline
Primary particle & ~~~~Energy & ~~~~Number of Showers \\\hline
$\gamma$-ray  &  ~~~100 GeV  & 10000 \\
              &  ~~~250 GeV  &  ~~7000 \\
              &  ~~~500 GeV  &  ~~5000 \\
              &  ~~~~~~~1 TeV    &  ~~2000 \\
              &  ~~~~~~~2 TeV    &  ~~1000 \\
              &  ~~~~~~~5 TeV    &  ~~~400 \\
  && \\
 Proton       &  ~~~250 GeV  & 10000 \\
              &  ~~~500 GeV  &  ~~8000 \\
              &  ~~~~~~~1 TeV    &  ~~5000 \\
              &  ~~~~~~~2 TeV    &  ~~2000 \\
              &  ~~~~~~~5 TeV    &  ~~~800 \\
 && \\
 Iron         &  ~~~~~~~5 TeV  & ~~5000 \\
              &  ~~~~~10 TeV  &  ~~2000 \\
              &  ~~~~~50 TeV    &  ~~~500 \\
              &  ~~~100 TeV    & ~~~100 \\
            
\hline
\end{tabular}
\end{table}
\end{center}
\vspace{-0.5cm}
The range of energies of the primaries used for the work corresponds to the 
typical ACT based observational energies in terms of their equivalent Cherenkov
photon yields. Using the HAGAR Experiment \cite{Chitnis} at Hanle as the 
reference, we have 
taken the observational height as 4270 m with the longitude: 78$^o$ 57$^\prime$
51$^{\prime\prime}$ E, the latitude: 32$^o$ 46$^\prime$ 46$^{\prime\prime}$ N,
and the magnetic field: H = 32.94 $\mu$T, Z = 38.29 $\mu$T. We have considered 
a flat 
horizontal detector array having 25 telescopes in the East-West direction with 
25 m separation between two consecutive detectors, and 25 telescopes in the 
North-South direction with 20 m separation between two consecutive detectors. 
This dimension of detector array is considered by taking into account the range
of energies and zenith angles at which showers were generated as shown Table 
\ref{tab1}. The shower core is assumed to coincide with the centre of the 
detector array. The individual detectors have an effective collection area of 
9 m$^{2}$ and the photons are counted in the steps where they are emitted with 
the emission angle independent of wavelength. Further, the wavelength window 
of Cherenkov photons is set at 200 nm to 600 nm for minimization of background 
noise and at the same time for collecting sufficient number of Cherenkov 
photons by detectors. The variable bunch size option of Cherenkov photon is set to '5' which is optimized for reduction of data size without losing the useful 
information. The parameter STEPFC is set at 0.1 and the energy cut-offs of 
kinetic energy for hadrons, muons, electrons and photons are chosen as 3.0 GeV,
3.0 GeV, 0.003 GeV and 0.003 GeV respectively, which are reasonable values for 
not to eliminate those parent particles which might decay to secondaries under 
investigation. The Linsley's parametrized US standard atmospheric model 
\citep{US} has been used here among the different atmospheric models available 
in CORSIKA, as we have not observed any significant impact of the different 
Atmospheric models on the distribution of Cherenkov photons from one of our 
earlier studies \cite{Das}.

\section{Analysis of the simulation and results}
\label{sec3}
To obtain the azimuthal distribution of the Cherenkov photons we have divided 
the detector array into four quadrants: with (+x,+y) coordinates as the first 
quadrant in the anticlockwise direction, (-x,+y) as the second quadrant, 
(-x,-y) as the third quadrant and (+x,-y) as the fourth quadrant. The detectors
in the first quadrant collect the photons emitted within 0 to 90 degree, 
detectors in the second quadrant within 90 to 180 degree, detectors in the 
third quadrant within 180 to 270 degree and the detectors in the fourth 
quadrant within 270 to 360 degree in the azimuthal plane. The Cherenkov photons
detected by all the detectors within a certain azimuthal angle bin are added to 
get the total Cherenkov photons per shower within that azimuthal angle bin. 
Repeating the same for all the showers we finally obtain the average value of
Cherenkov photons per shower for a given azimuthal angle bin. These data are 
then represented as histograms of suitable azimuthal angle bins. To 
understand the pattern of distribution of Cherenkov photons in the azimuthal 
plane, we also collected the number of electrons and positrons for the same
azimuthal angle bin following the same procedure as stated above. Moreover,  
the electron and positron asymmetry ($A_{S}$) has been calculated by using the 
formula given as \cite{Goswami} $$A_{S} = \dfrac{N_{e}-N_{p}}{N_{e}+N_{p}},$$ 
where $N_{e}$ and $N_{p}$ are the photon counts corresponding to the electron 
peak and the positron peak respectively of the Cherenkov photon's distribution 
histogram. The analysis has been carried out on the root platform \cite{Root} 
by using C++ programs. The results of this works are discussed in the following sub sections:

\subsection{Azimuthal distribution of Cherenkov photons}
\begin{figure}[hbt]
\centerline
\centerline{
\includegraphics[scale = 0.29]{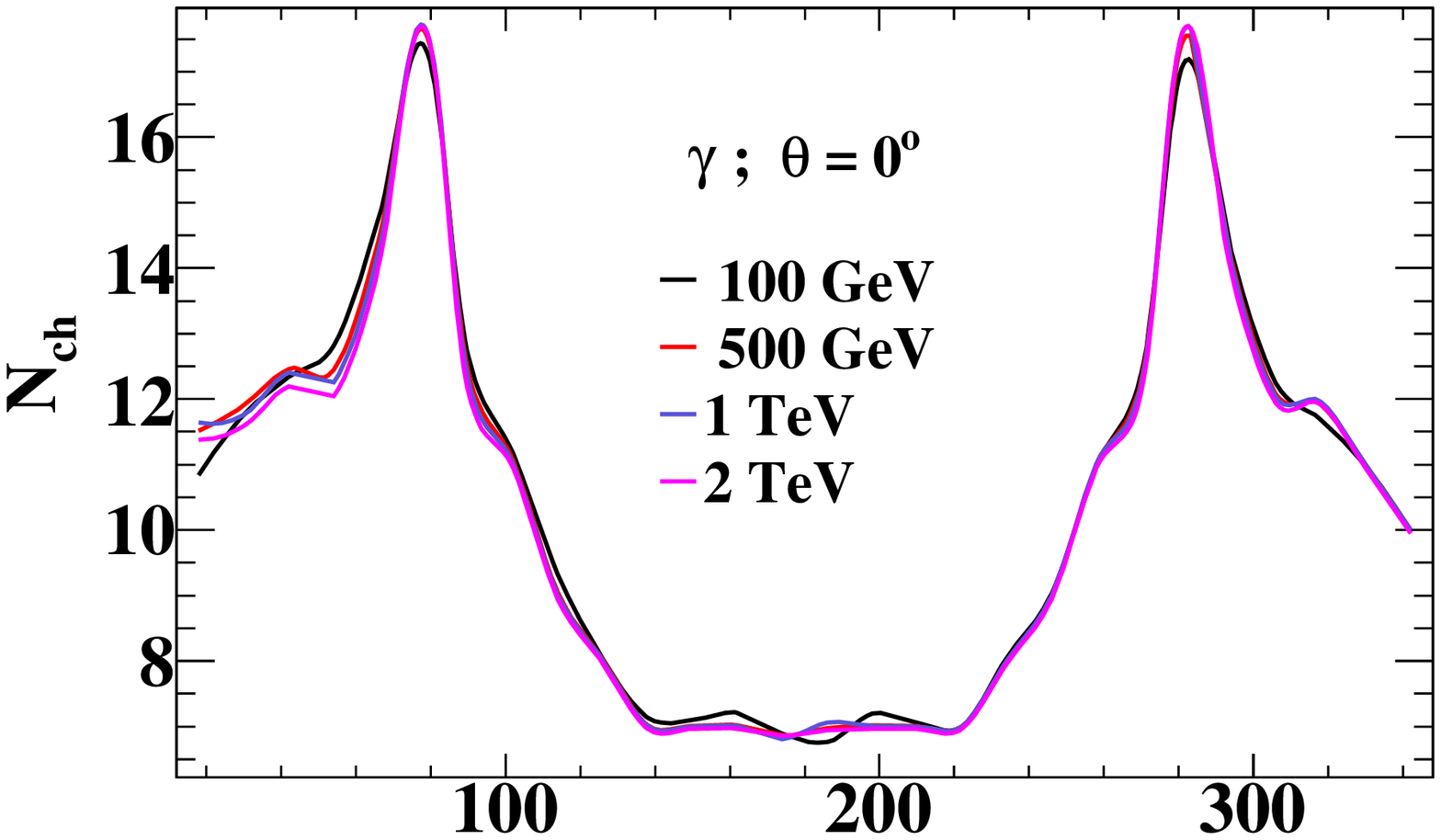}\hspace{-0.3cm}
\includegraphics[scale = 0.29]{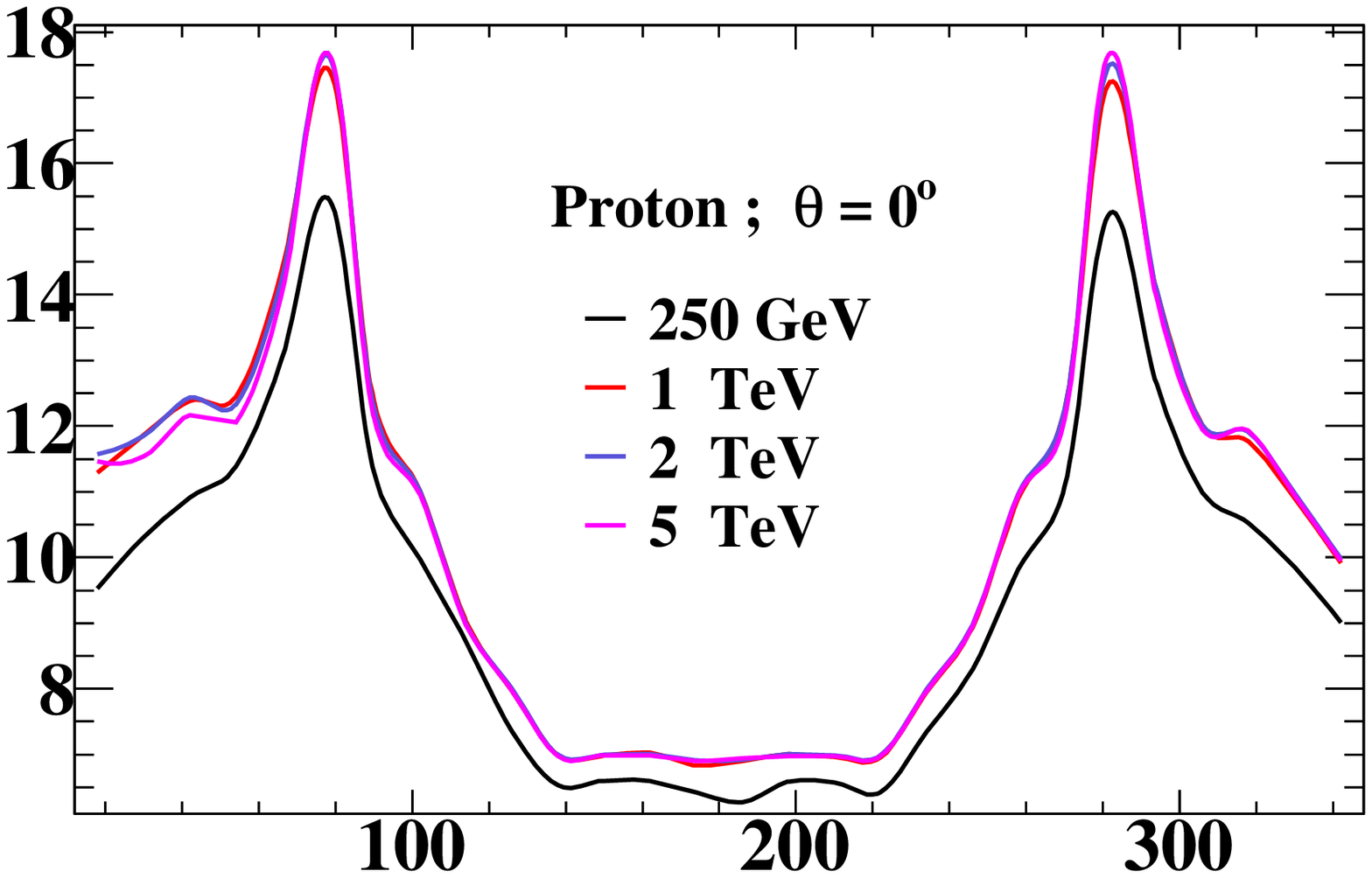}\hspace{-0.3cm}
\includegraphics[scale = 0.29]{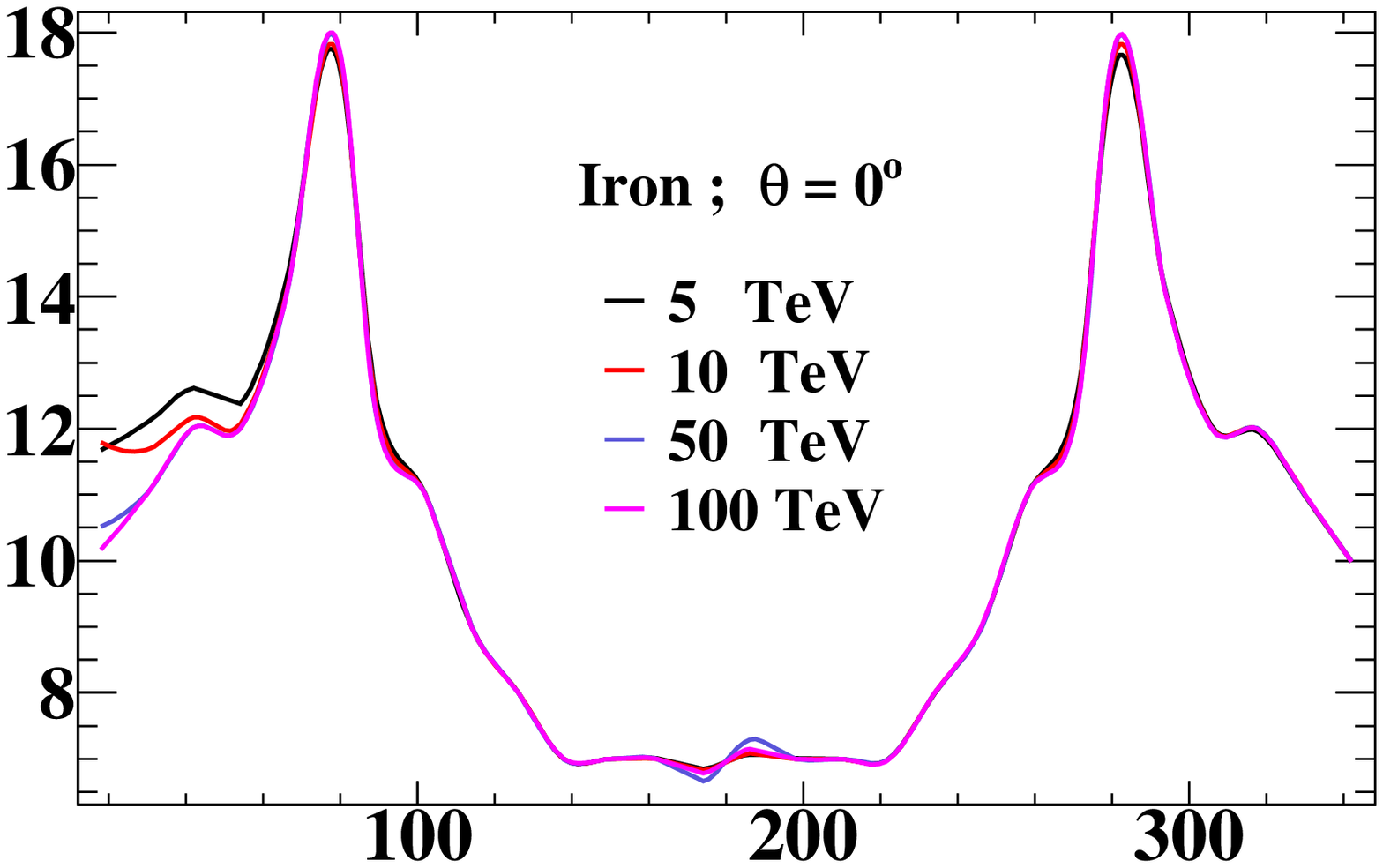}}
\centerline{
\includegraphics[scale = 0.29]{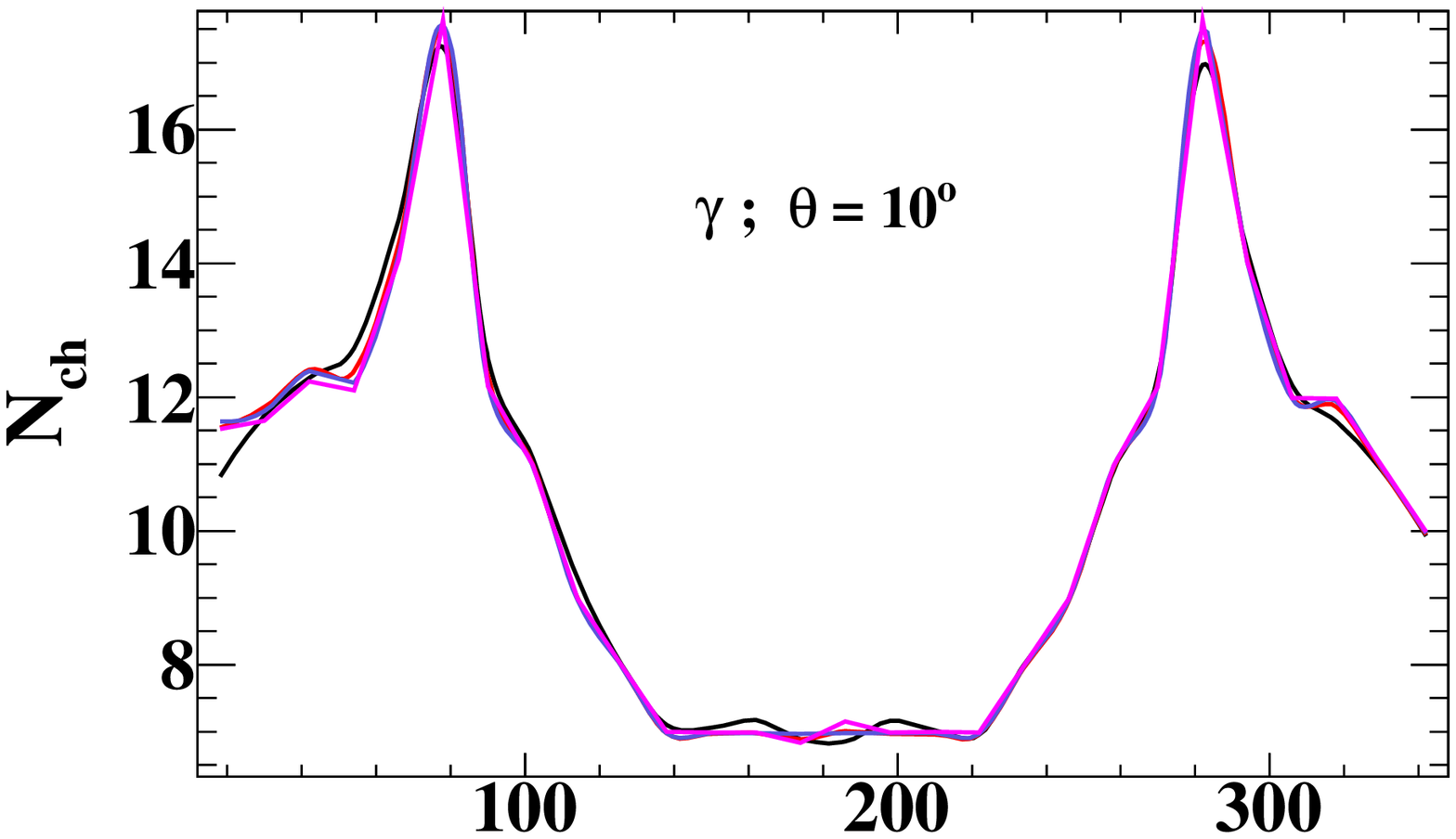}\hspace{-0.3cm}
\includegraphics[scale = 0.29]{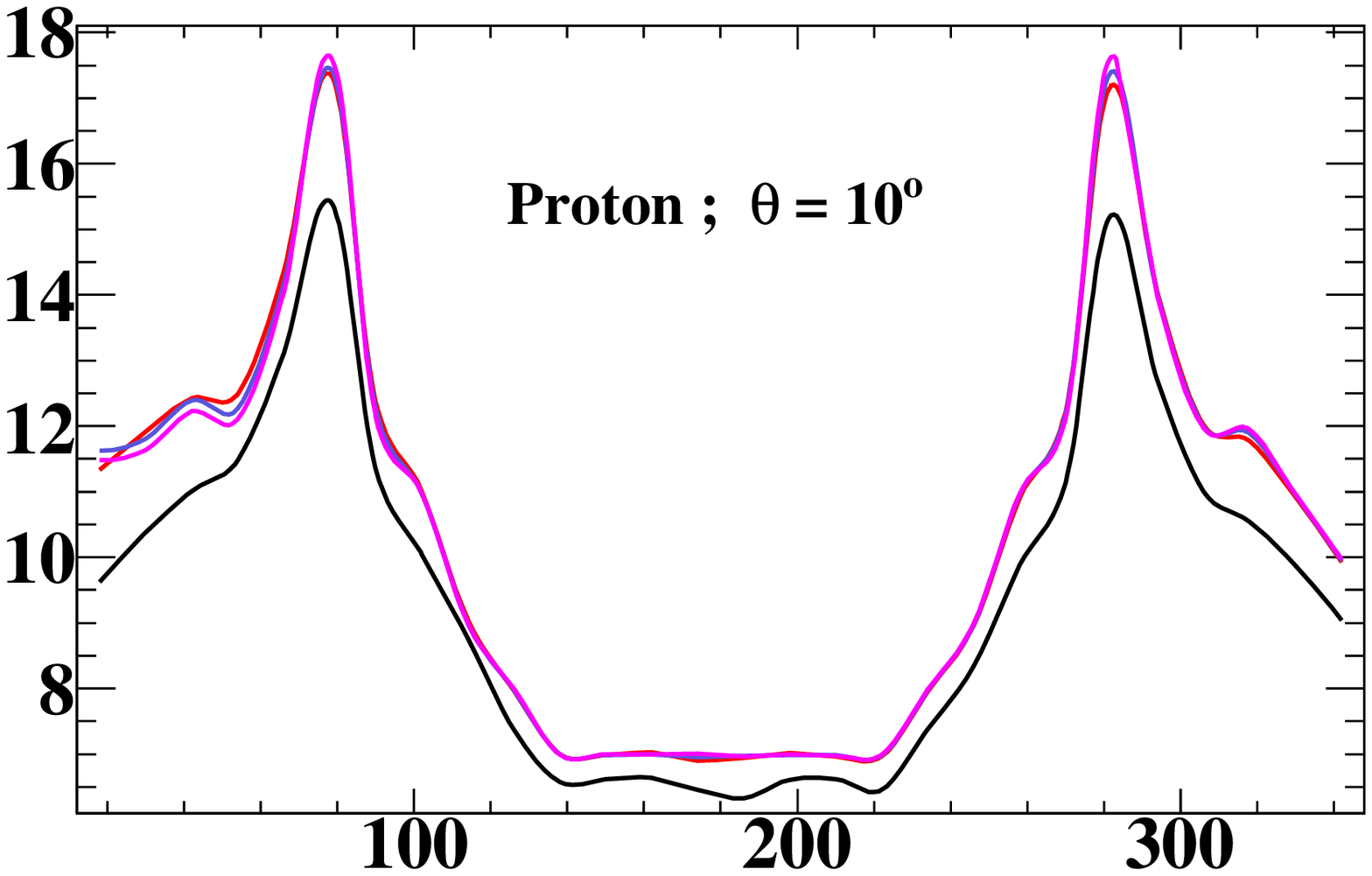}\hspace{-0.3cm}
\includegraphics[scale = 0.29]{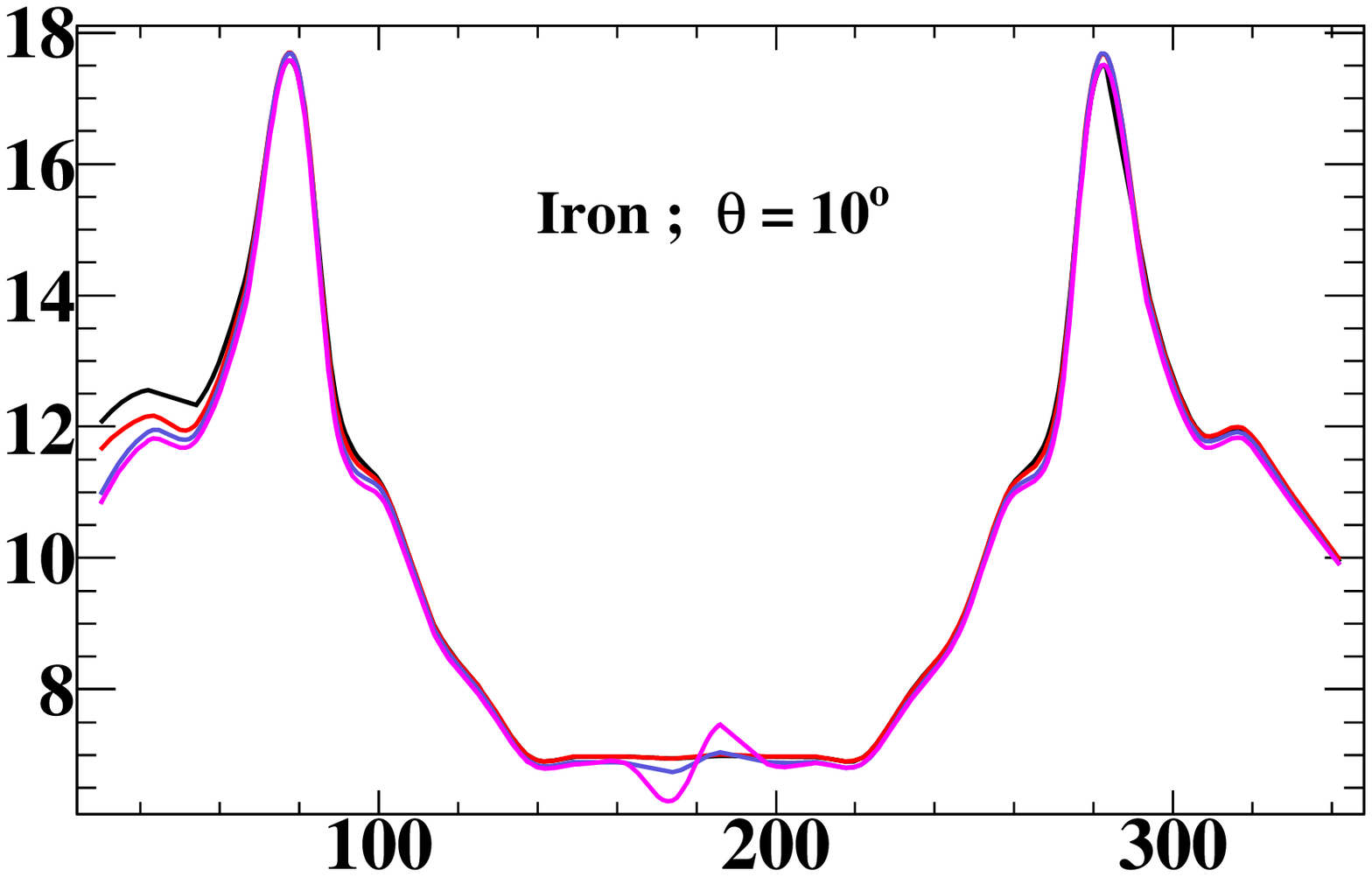}}
\centerline{
\includegraphics[scale = 0.29]{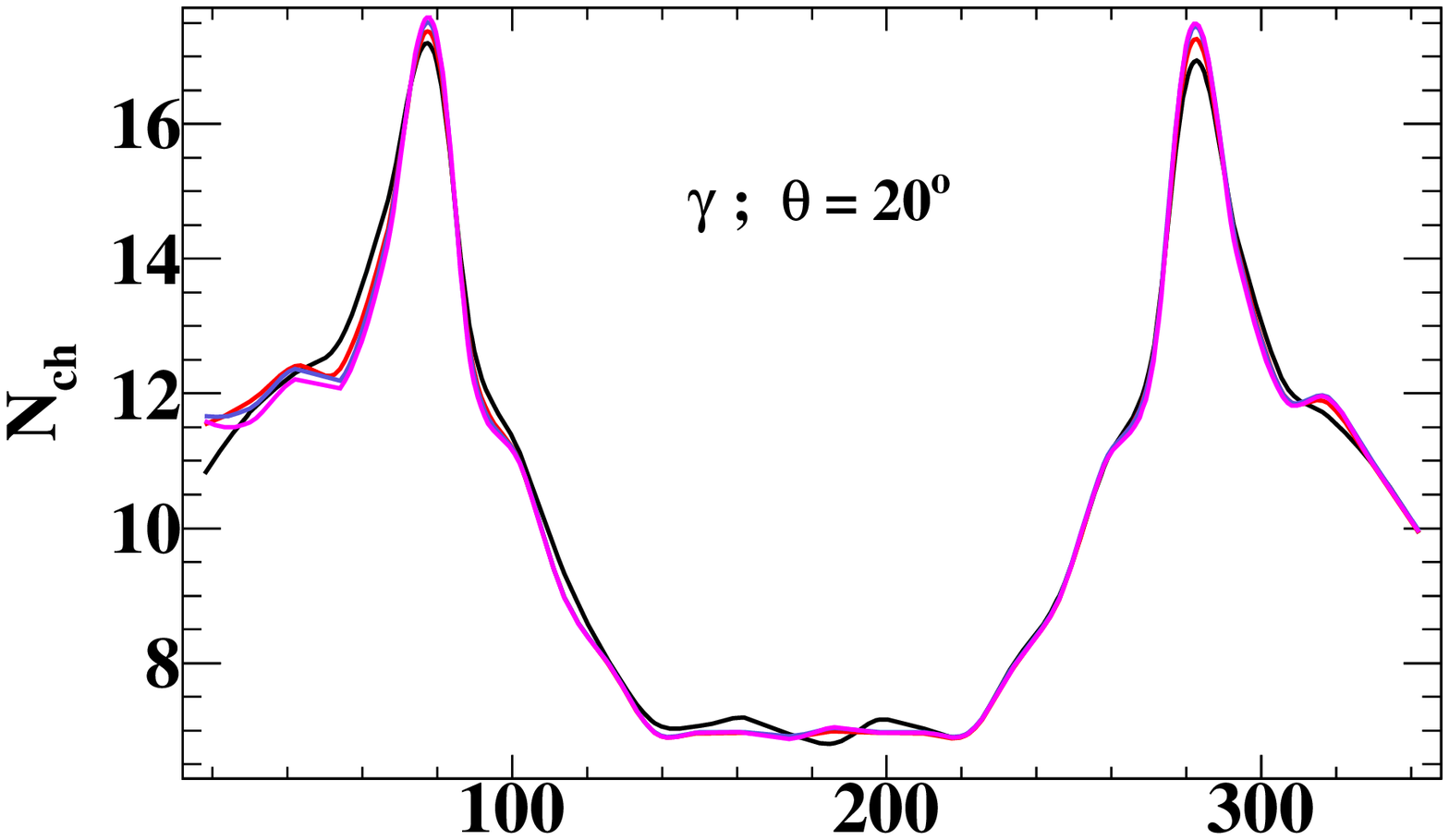}\hspace{-0.3cm}
\includegraphics[scale = 0.29]{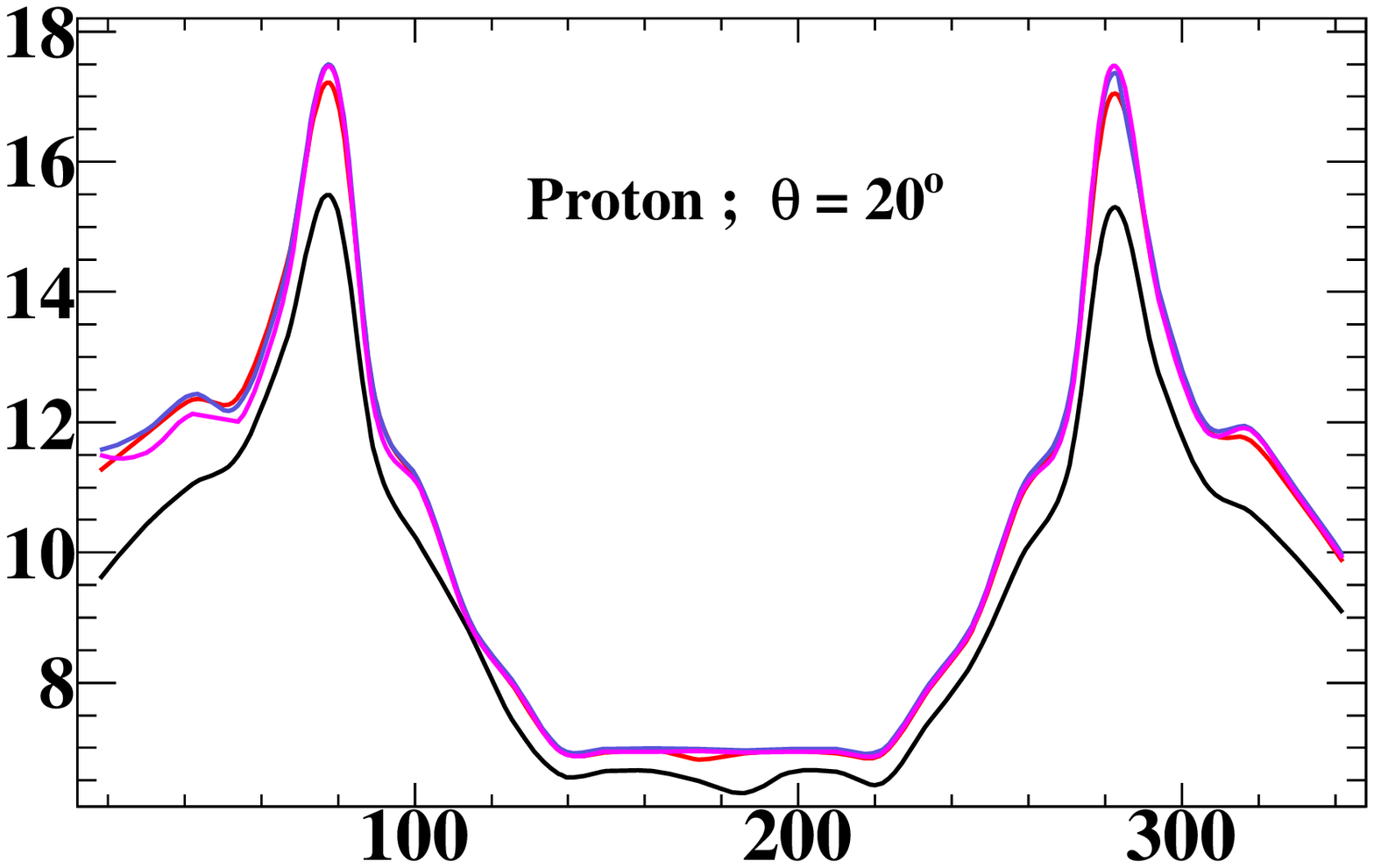}\hspace{-0.3cm}
\includegraphics[scale = 0.29]{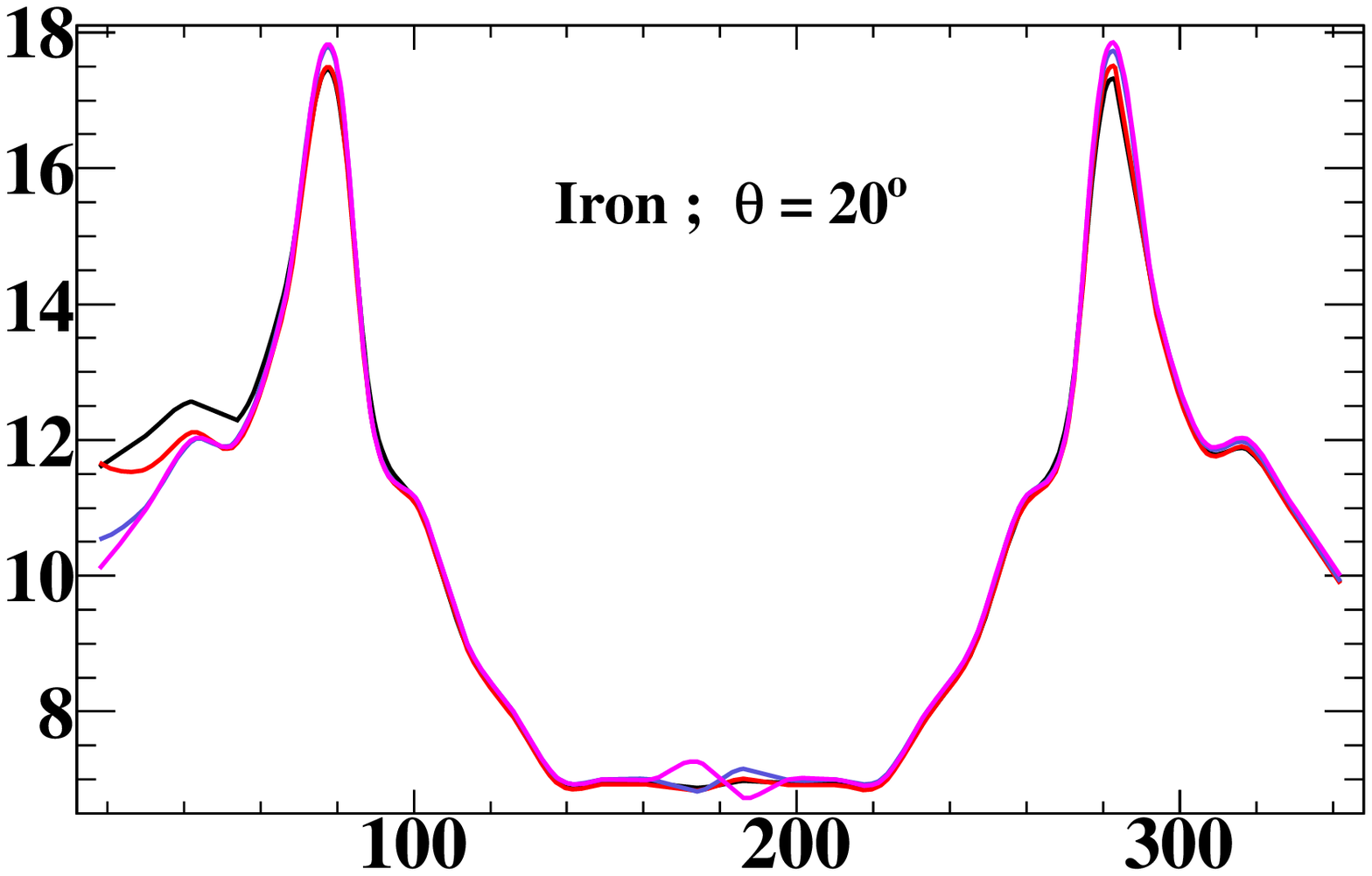}}
\centerline{
\includegraphics[scale = 0.29]{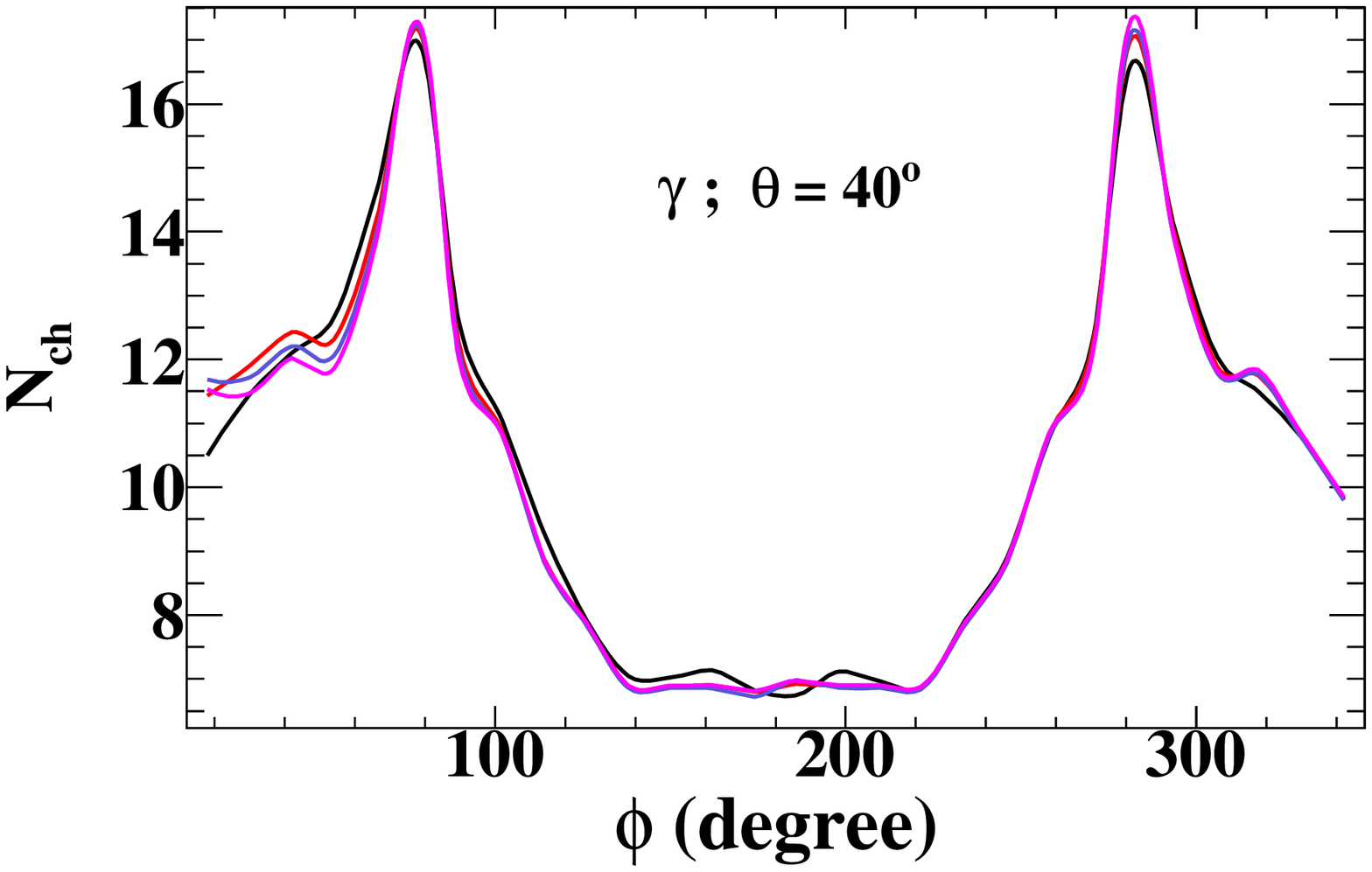}\hspace{-0.3cm}
\includegraphics[scale = 0.29]{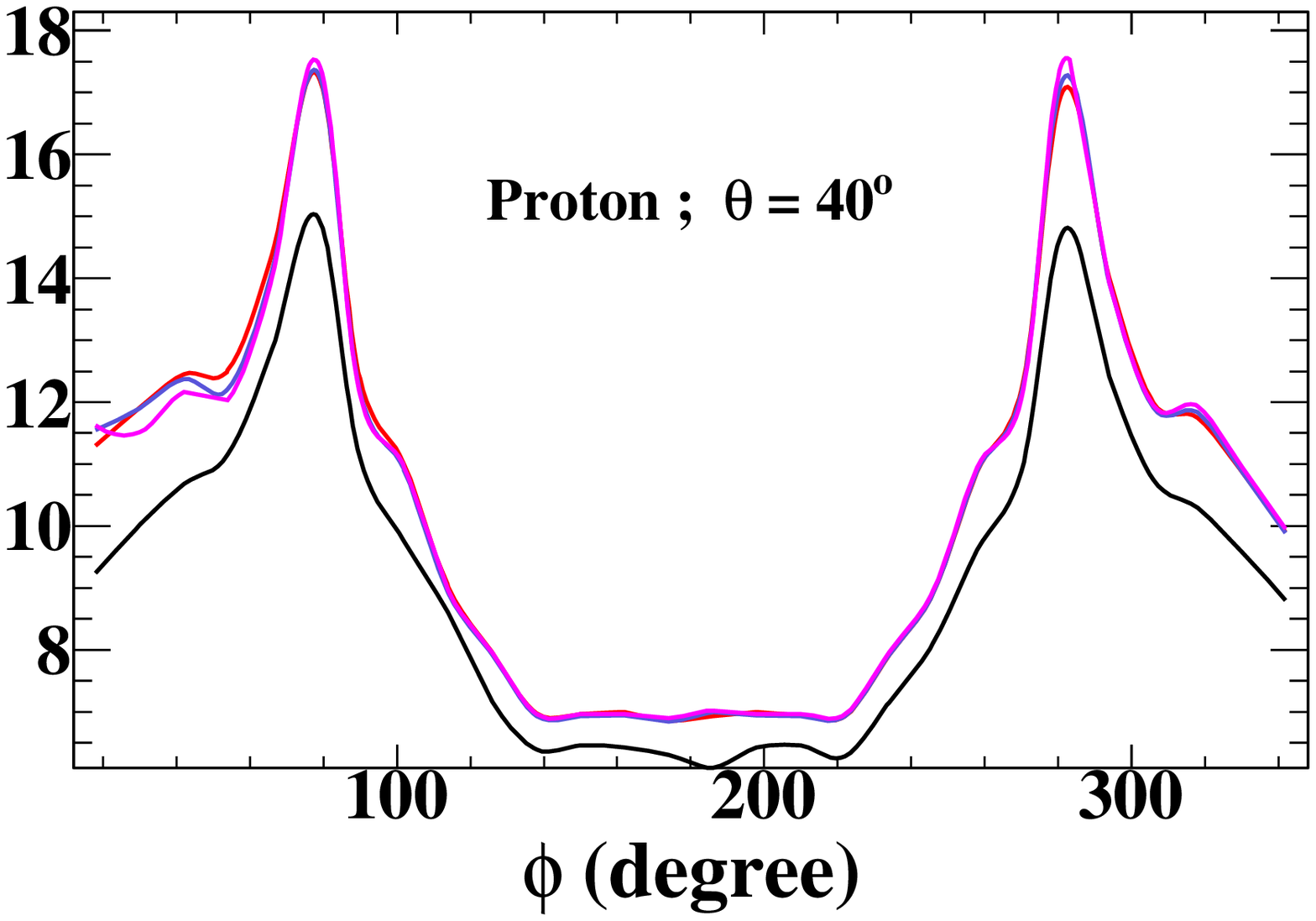}\hspace{-0.3cm}
\includegraphics[scale = 0.29]{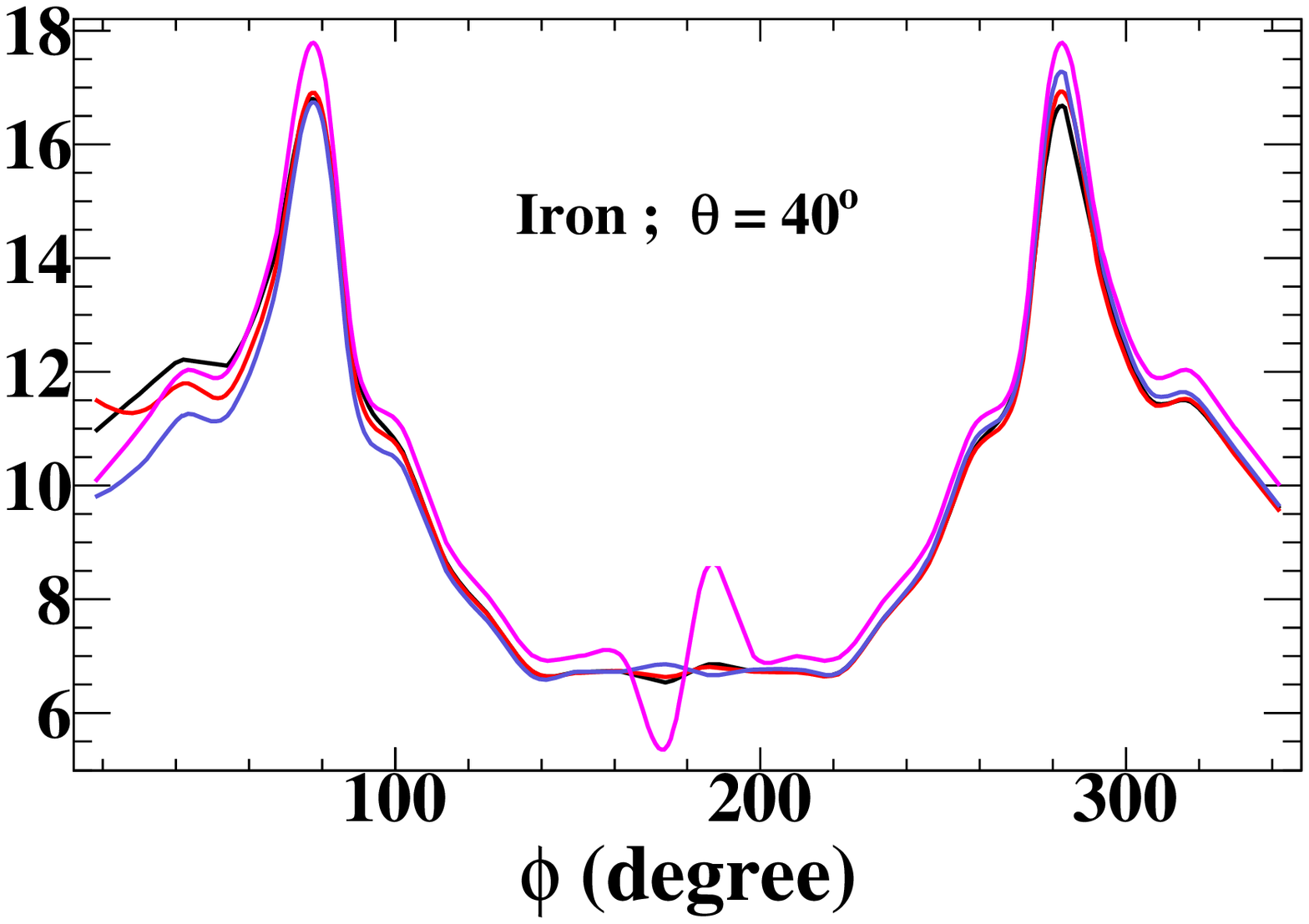}}
\caption{Smoothed histograms of Cherenkov photons with respect to azimuthal 
angle for  $\gamma$-ray, proton and iron primaries for different energies and 
at a fixed angle of incidence.}
\label{fig1}
\end{figure}

\begin{figure}[hbt]
\centerline
\centerline{
\includegraphics[scale = 0.29]{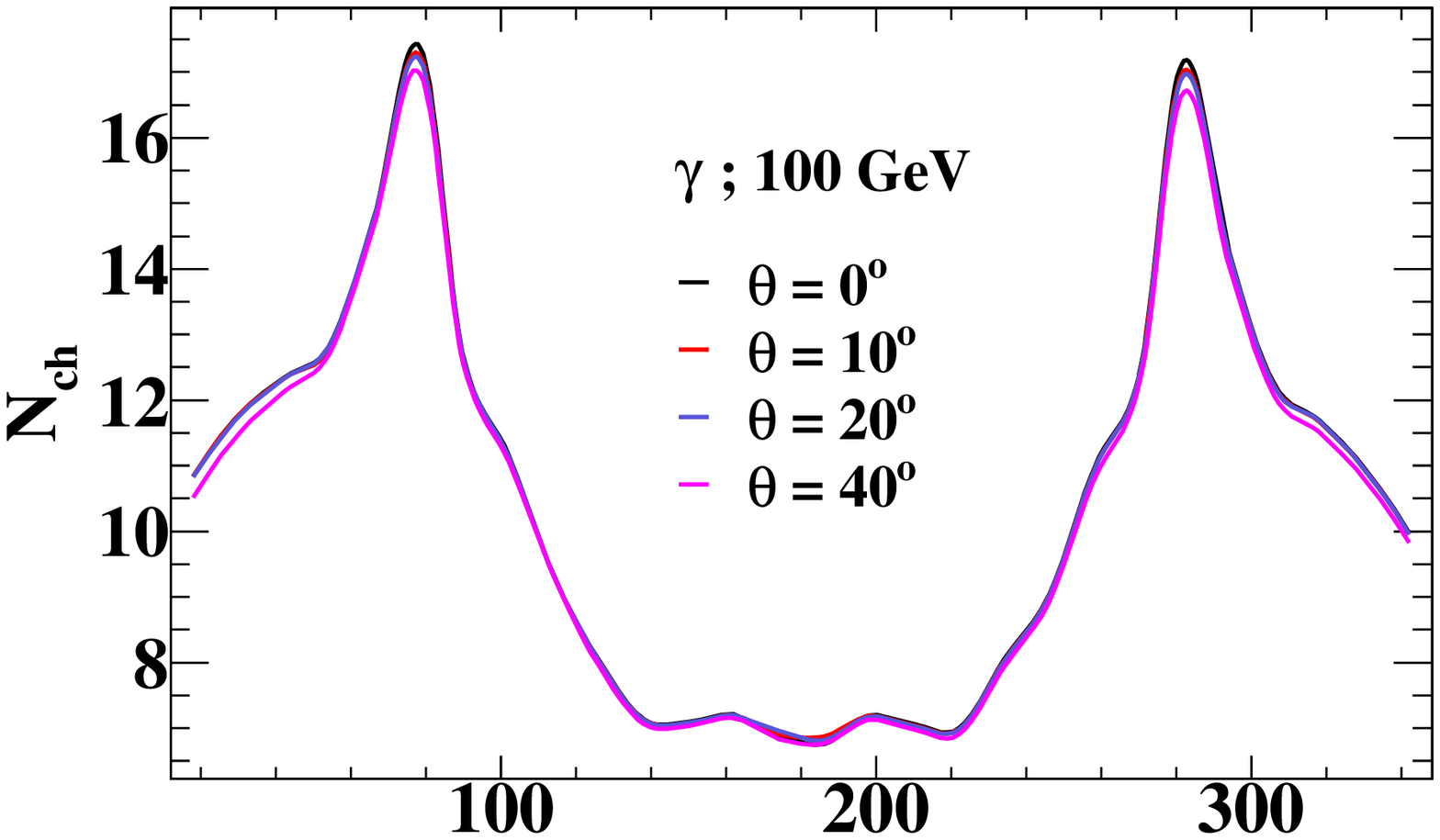}\hspace{-0.3cm}
\includegraphics[scale = 0.29]{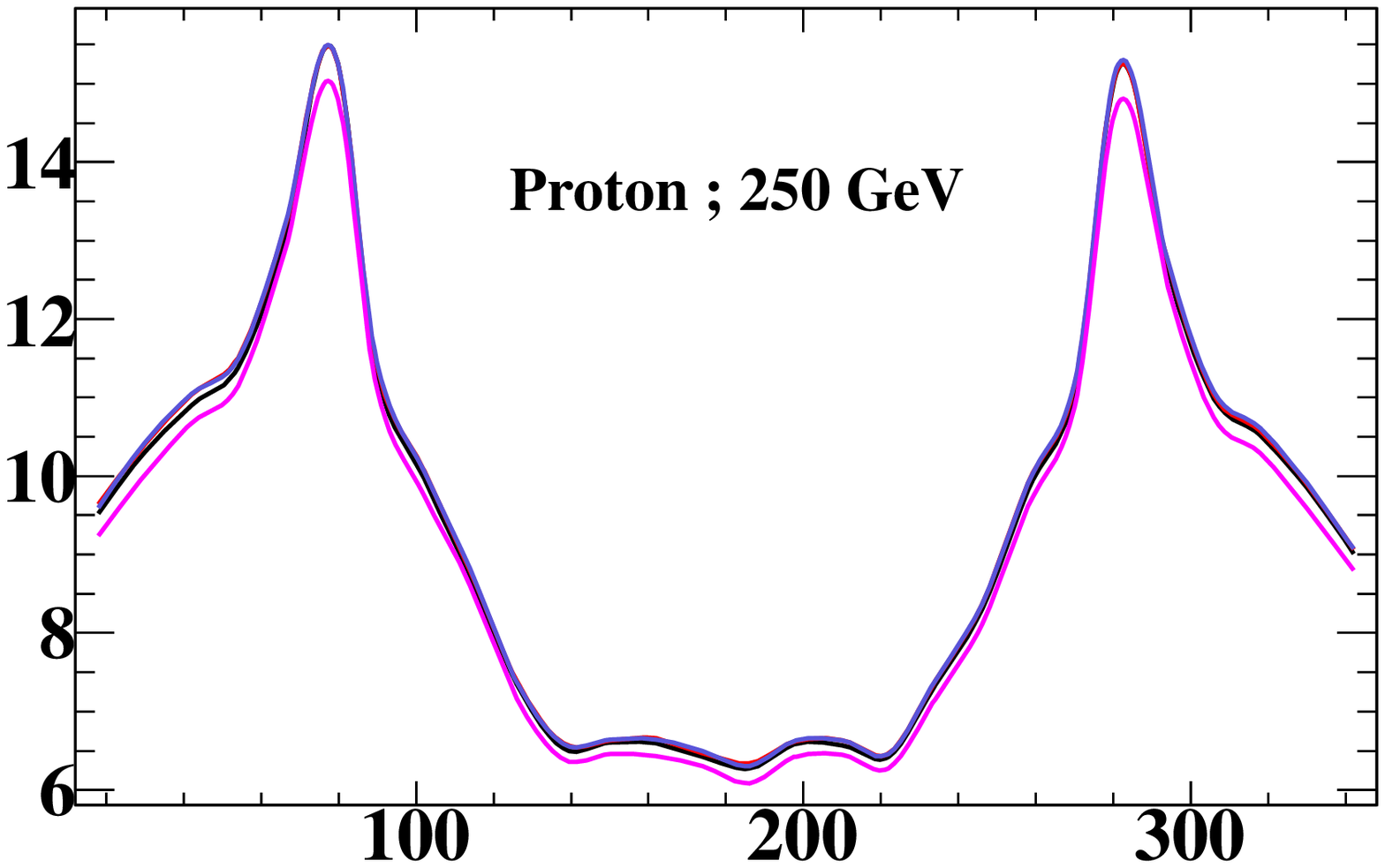}\hspace{-0.3cm}
\includegraphics[scale = 0.29]{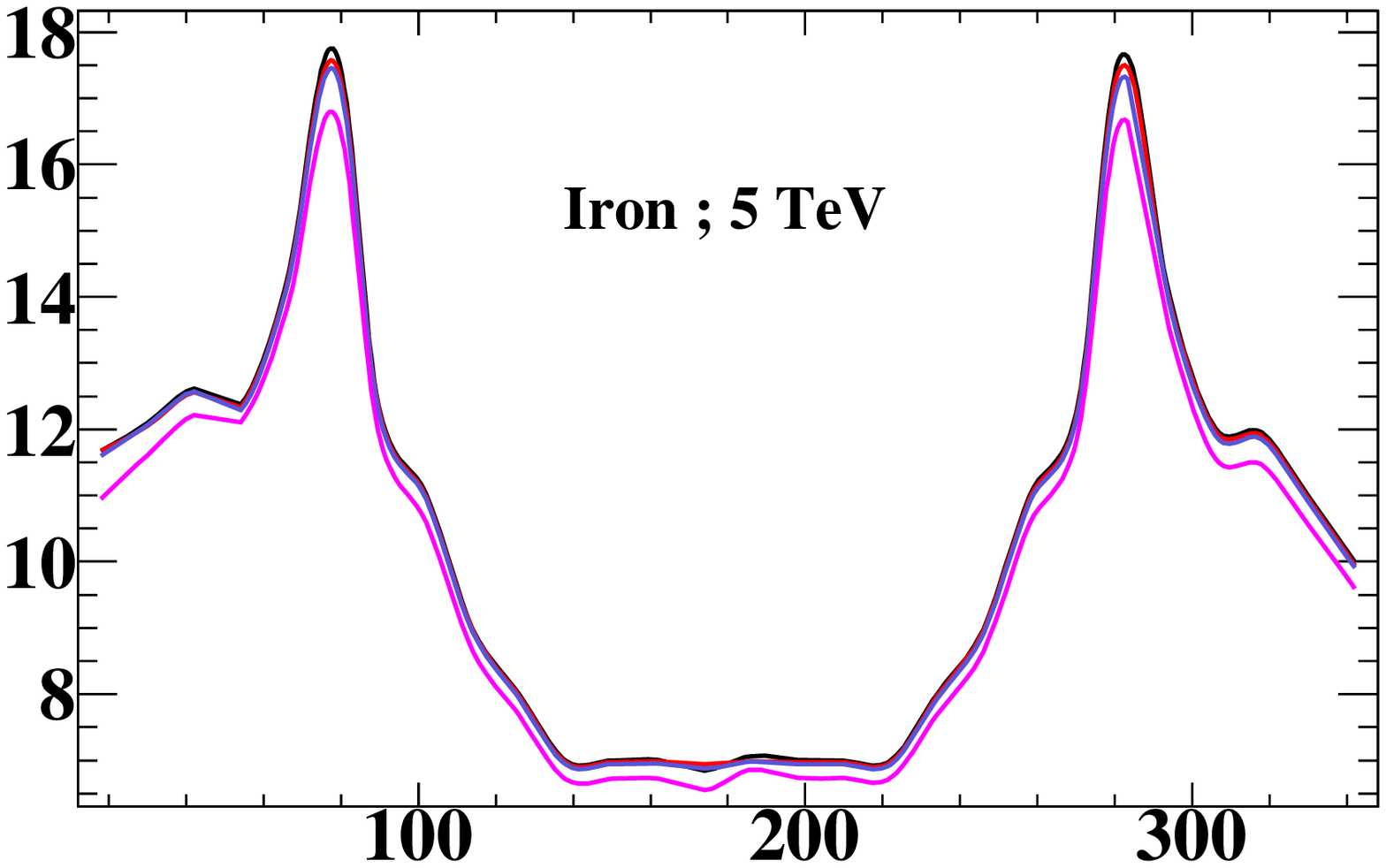}}
\centerline{
\includegraphics[scale = 0.29]{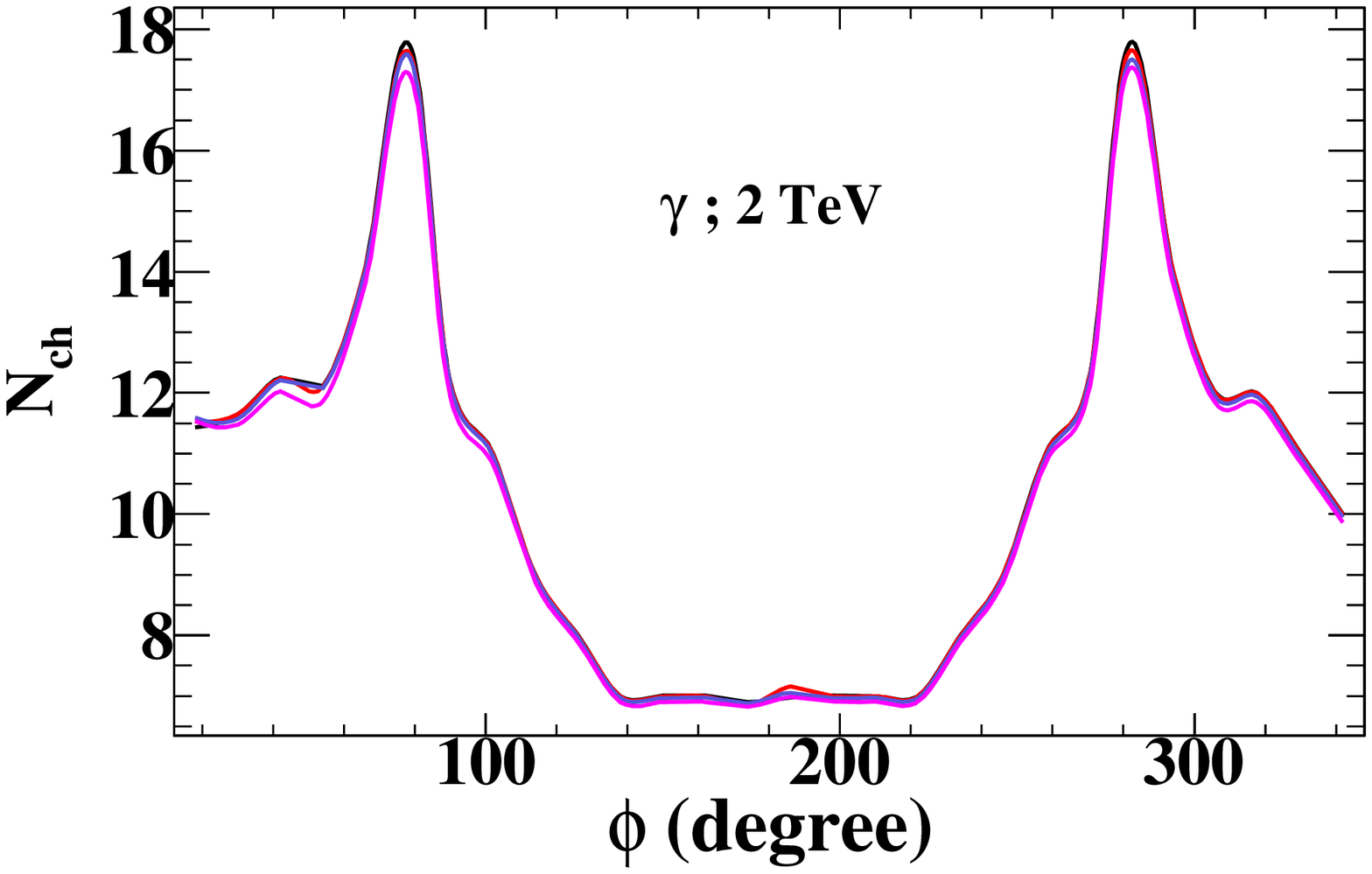}\hspace{-0.3cm}
\includegraphics[scale = 0.29]{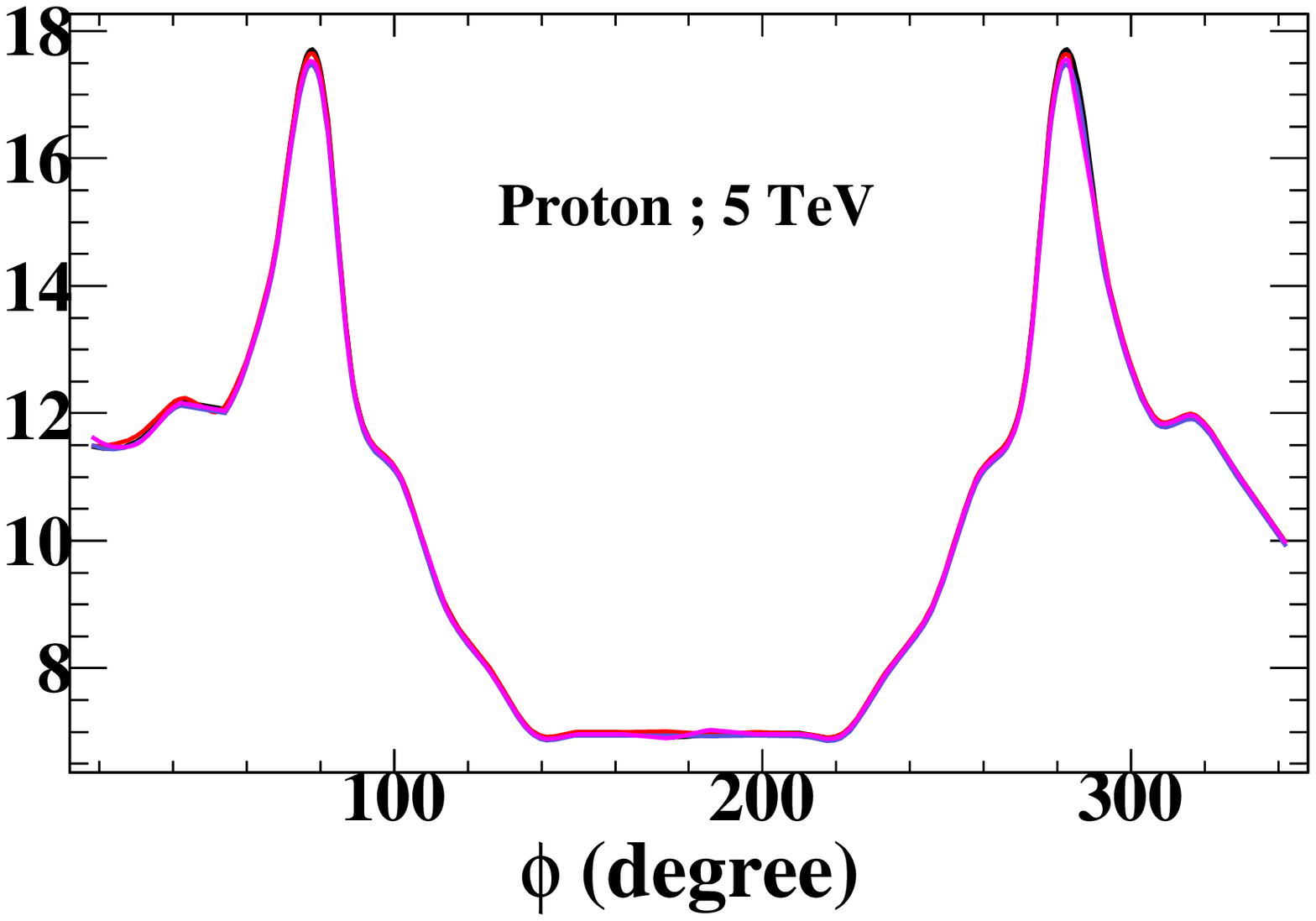}\hspace{-0.3cm}
\includegraphics[scale = 0.29]{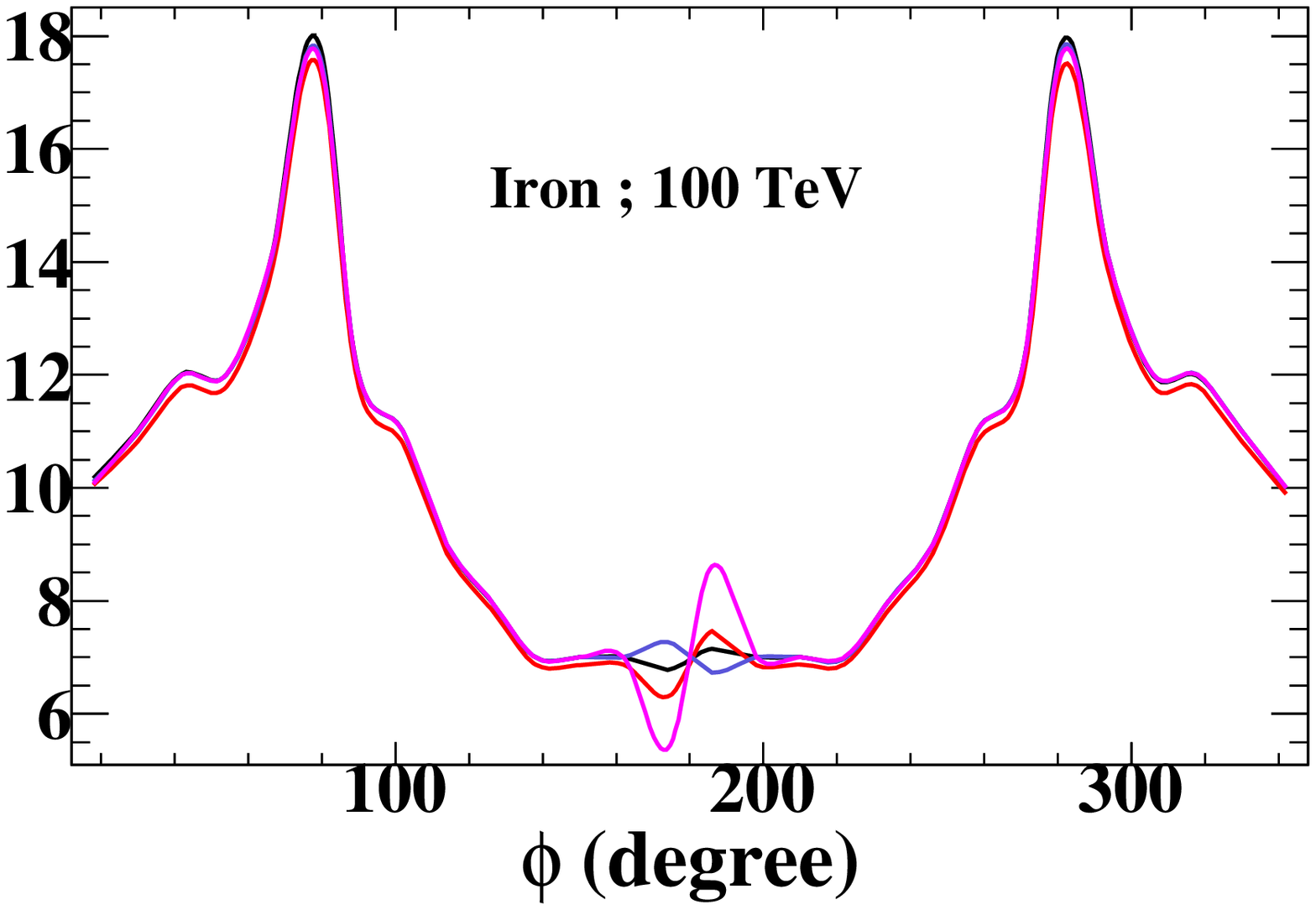}}
\caption{Smoothed histograms of Cherenkov photons with respect to azimuthal 
angle for  $\gamma$-ray, proton and iron primaries for different angle of 
incidence and at a fixed energy of the primary.}
\label{fig2}
\end{figure}

Fig.\ref{fig1} shows the smoothed histograms of Cherenkov photon counts as a 
function of the azimuthal angle initiated by the $\gamma$-ray, proton and iron 
primaries for different combinations of energies at a particular angle of 
incidence. In all the cases we can see the double peak distribution pattern 
with the first peak occurring in between 70$^{\circ}$ and 90$^{\circ}$, 
whereas the second peak occurring in between 270$^{\circ}$ and 290$^{\circ}$. 
The left peak occurs due to the Cherenkov photons produced by the electrons 
in EAS, whereas the right peak occurs because of the positrons. Actually, the
separation of electrons and positrons in EAS takes place over the azimuthal 
plane due to the opposite effect of earth's magnetic field on their charges. 
Thus the double peak distribution of Cherenkov photons in EAS over the 
azimuthal plane is a signature effect of the electron-positron separation over
the azimuthal plane due to the effect of earth's magnetic field on them. 
Fig.\ref{fig2} shows the similar azimuthal distribution patterns of Cherenkov 
photons initiated by the three primaries but for a fixed energy and variable 
angle of incidence. The distribution patterns do not show any major differences
between the $\gamma$-ray and hadron initiated showers. However, there is a small
difference in the overlapping zone near about 180$^{\circ}$ for the iron 
initiated showers. The prominence of this difference increases with increasing
energy and angle of incidence of the iron primary. Moreover, except for the
proton primaries the pattern of distributions appears to almost independent
of energy and angle of incidence. But the difference in distributions with
respect to energy is more sensitive than that with respect to angle of 
incidence, specially for the proton primary.  

\begin{figure}[hbt]
\centerline
\centerline{
\includegraphics[scale = 0.29]{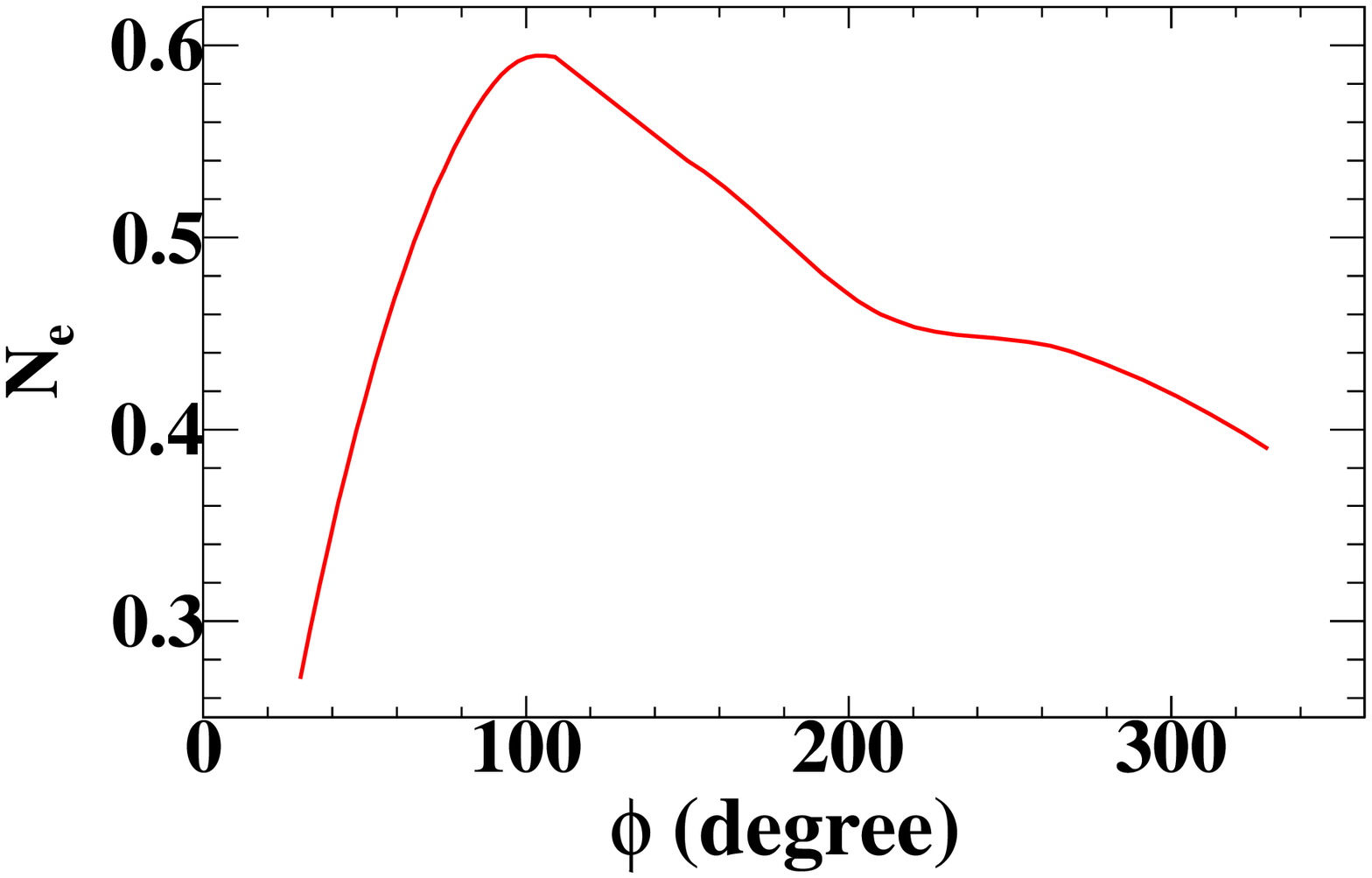} \hspace{0.5cm}
\includegraphics[scale = 0.29]{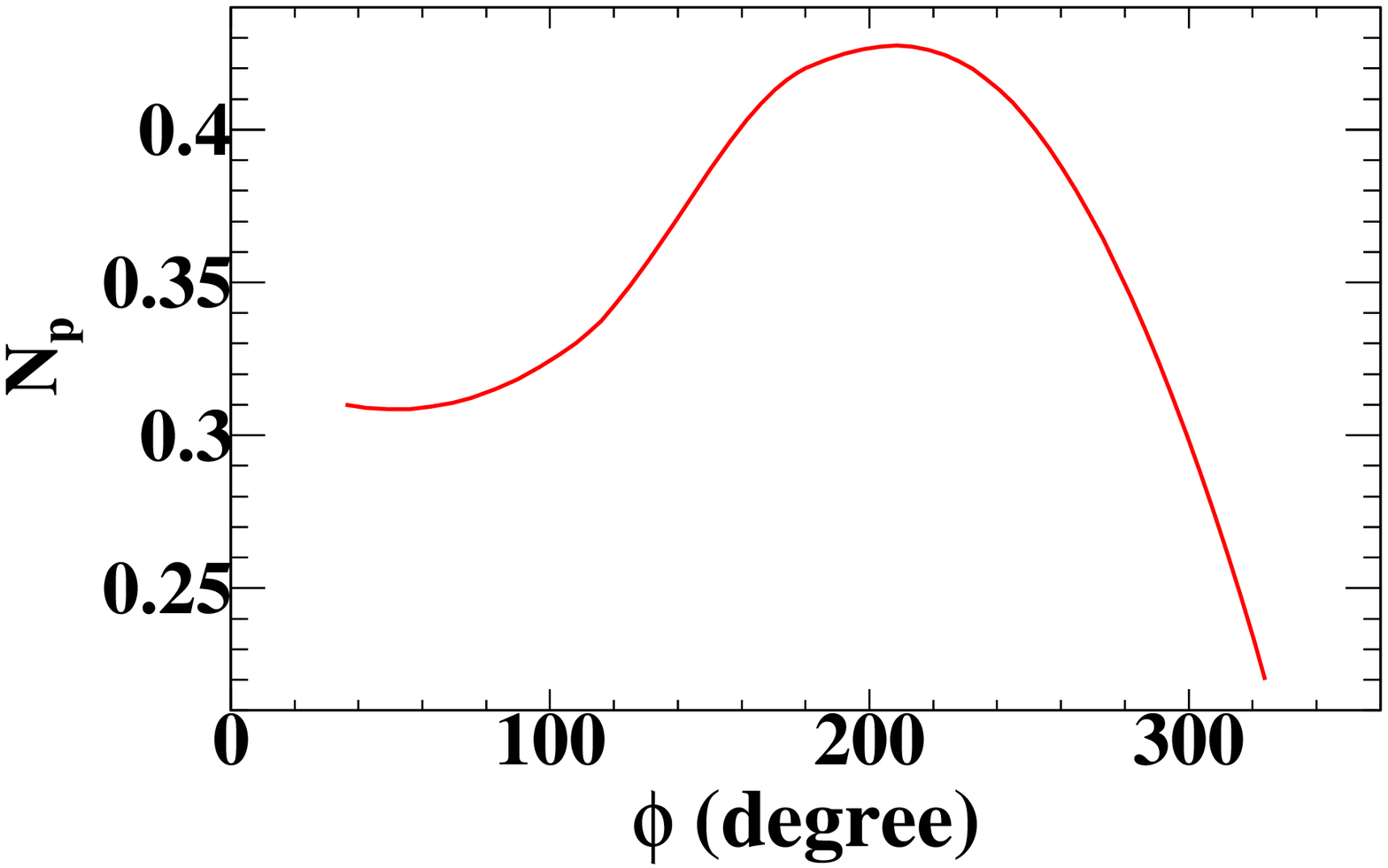}}
\caption{Azimuthal distributions of electrons and positrons initiated by 
vertically incident 1 TeV $\gamma$-rays.}
\label{fig3}
\end{figure}

\begin{figure}[hbt]
\centerline
\centerline{
\includegraphics[scale = 0.29]{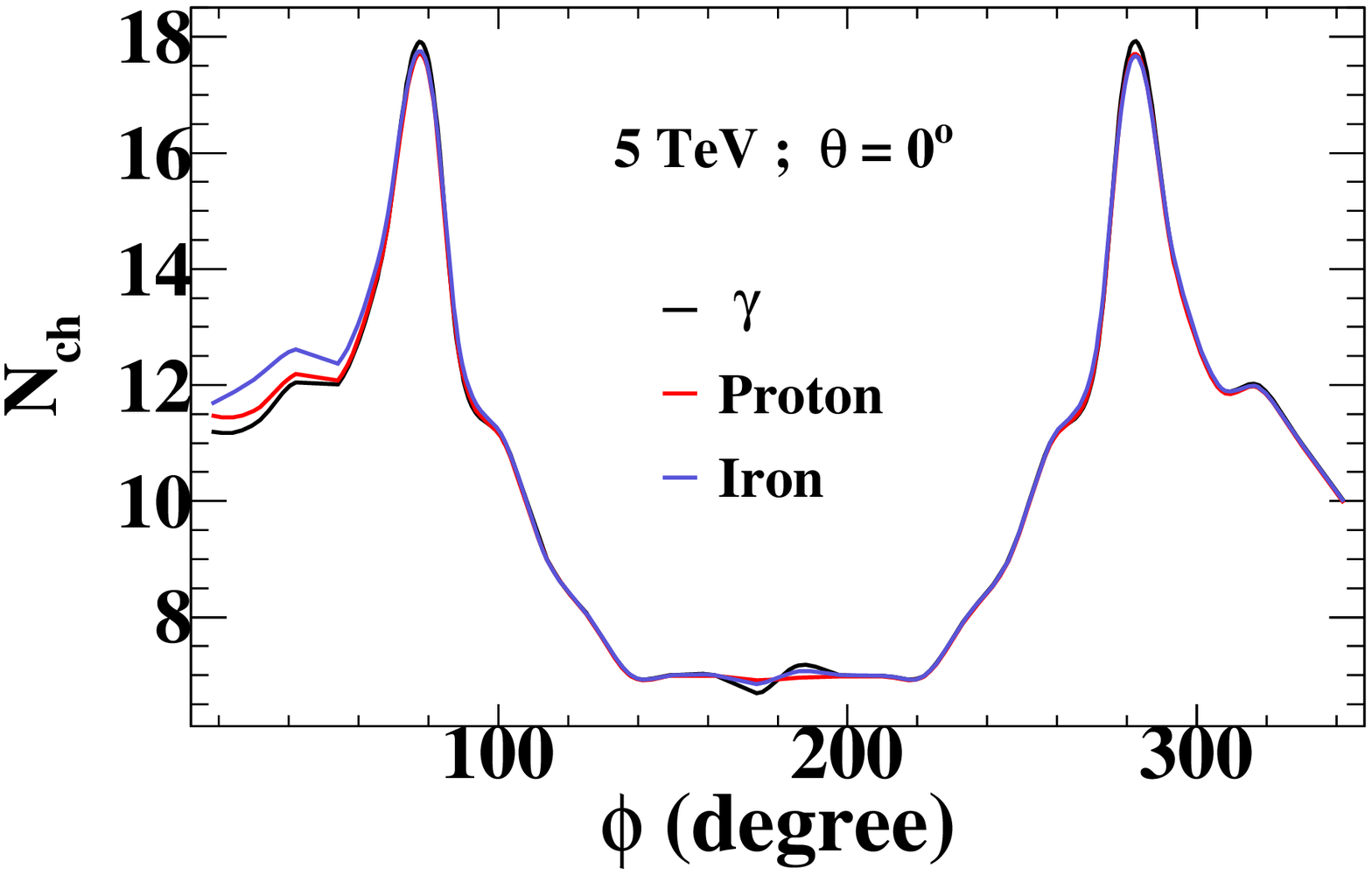}\hspace{-0.3cm}
\includegraphics[scale = 0.29]{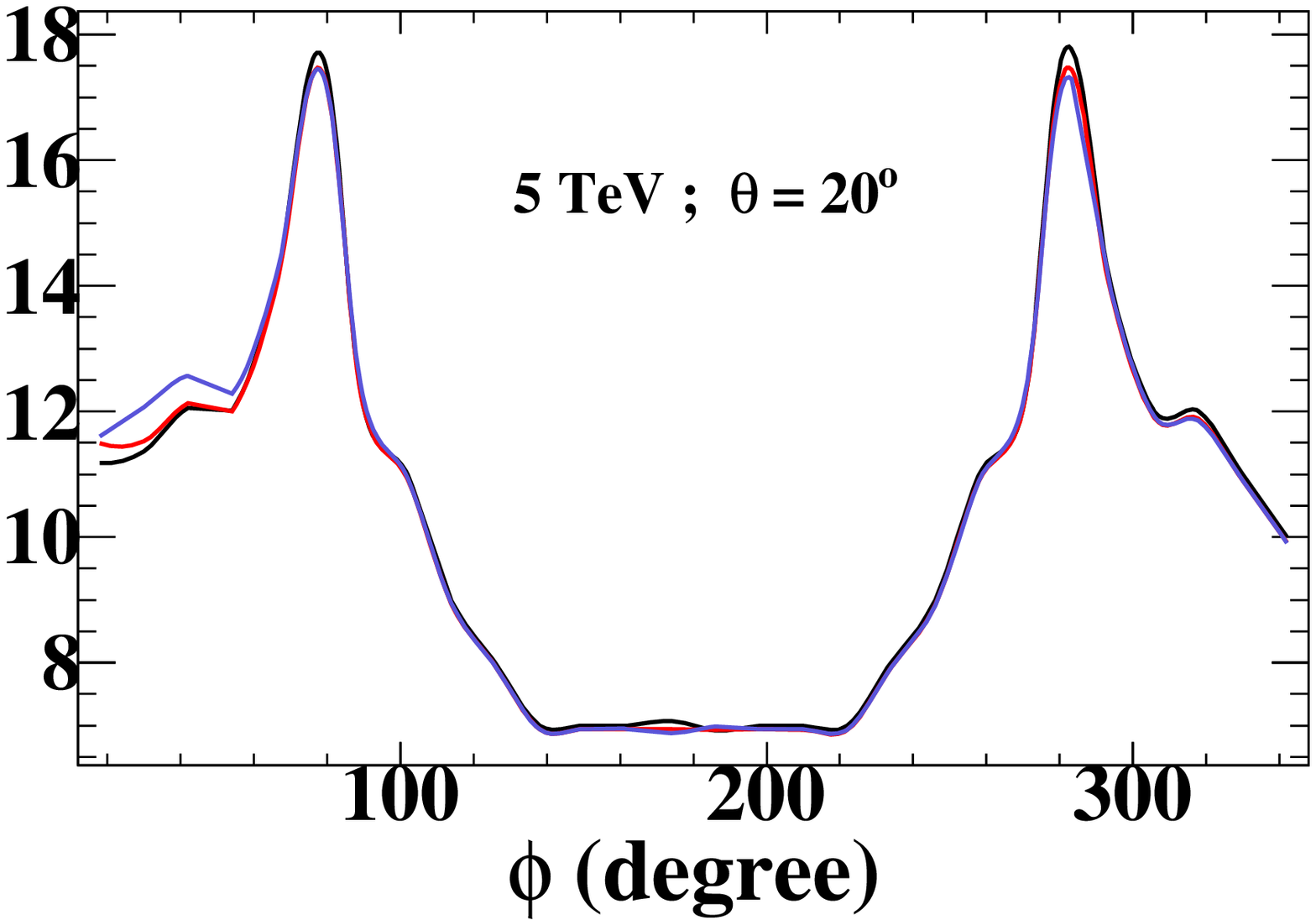}\hspace{-0.3cm}
\includegraphics[scale = 0.29]{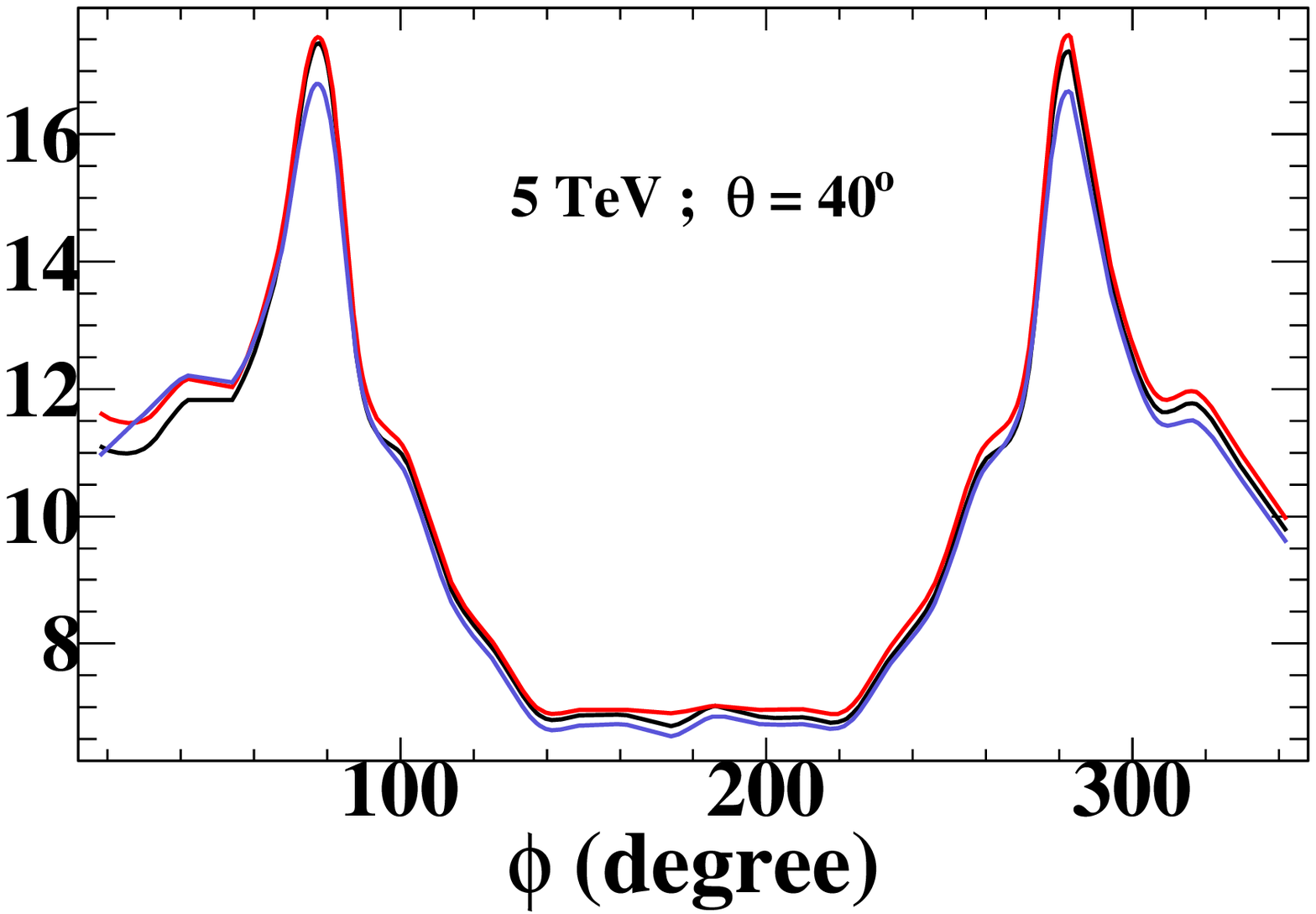}}
\caption{Smoothed histograms of Cherenkov photons with respect to azimuthal 
angle for  $\gamma$-ray, proton and iron primaries for same energy and same 
angle of incidence.}
\label{fig4}
\end{figure}

\begin{figure}[hbt]
\centerline
\centerline{
\includegraphics[scale = 0.29]{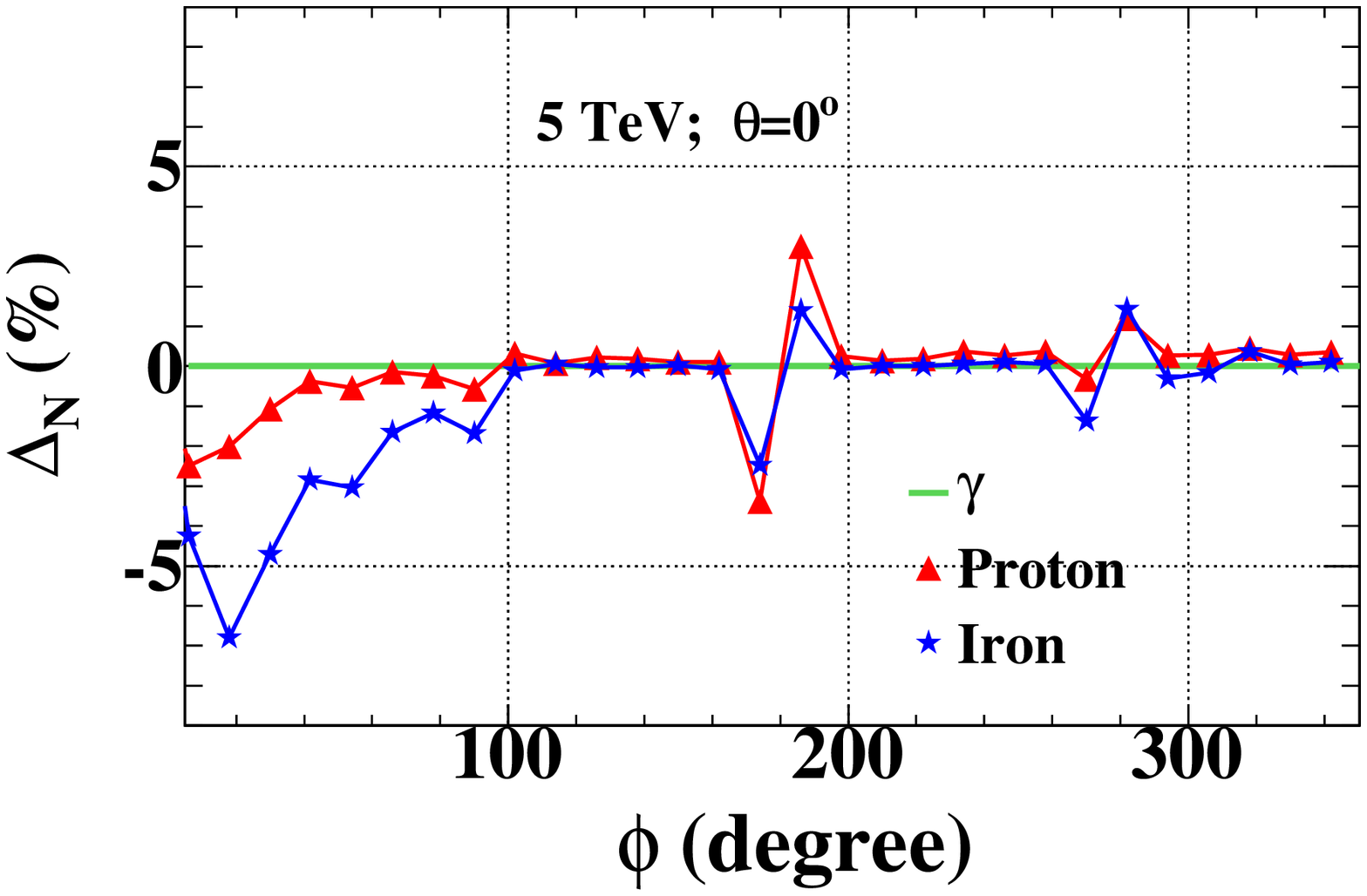}\hspace{-0.3cm}
\includegraphics[scale = 0.29]{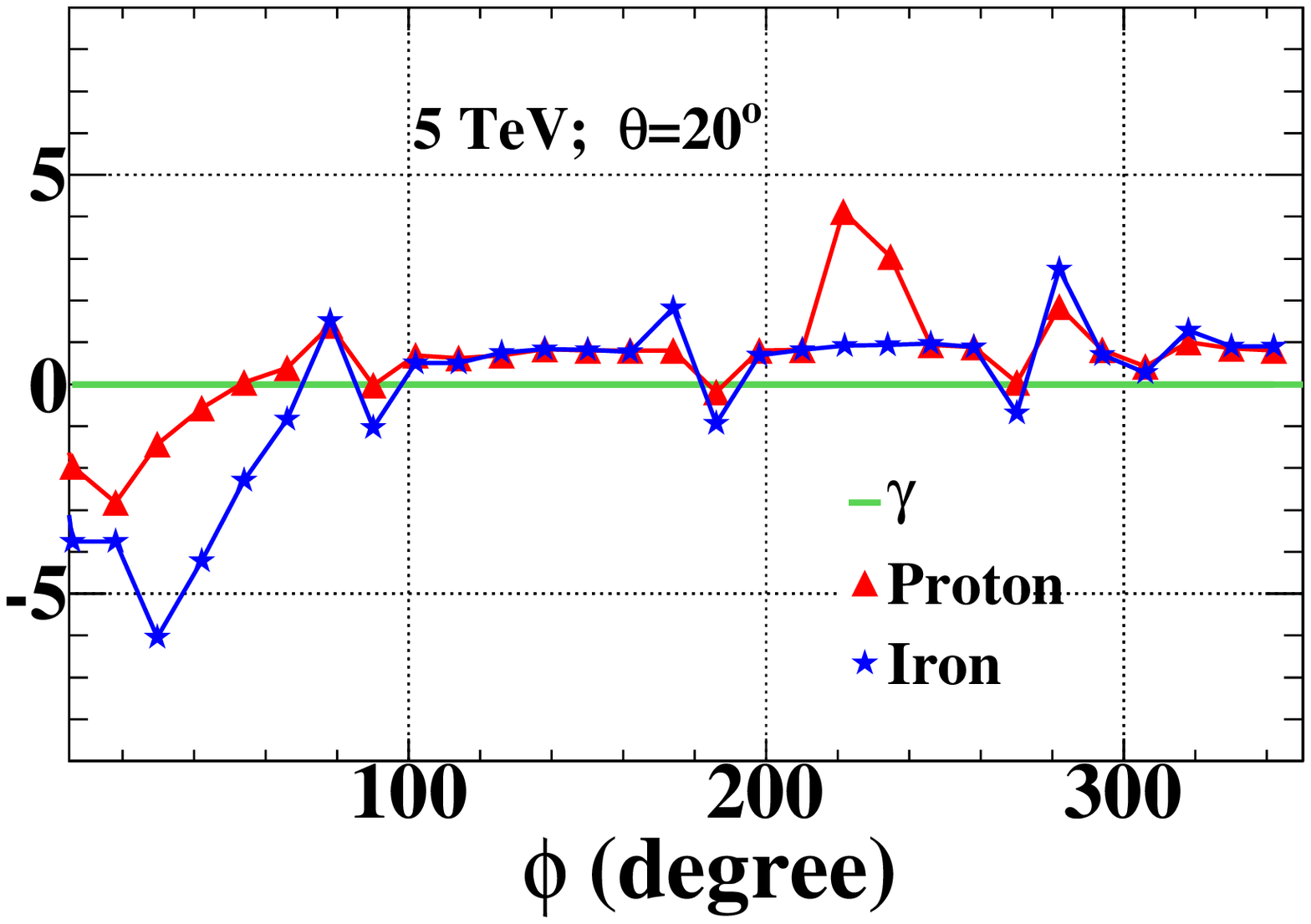}\hspace{-0.3cm}
\includegraphics[scale = 0.29]{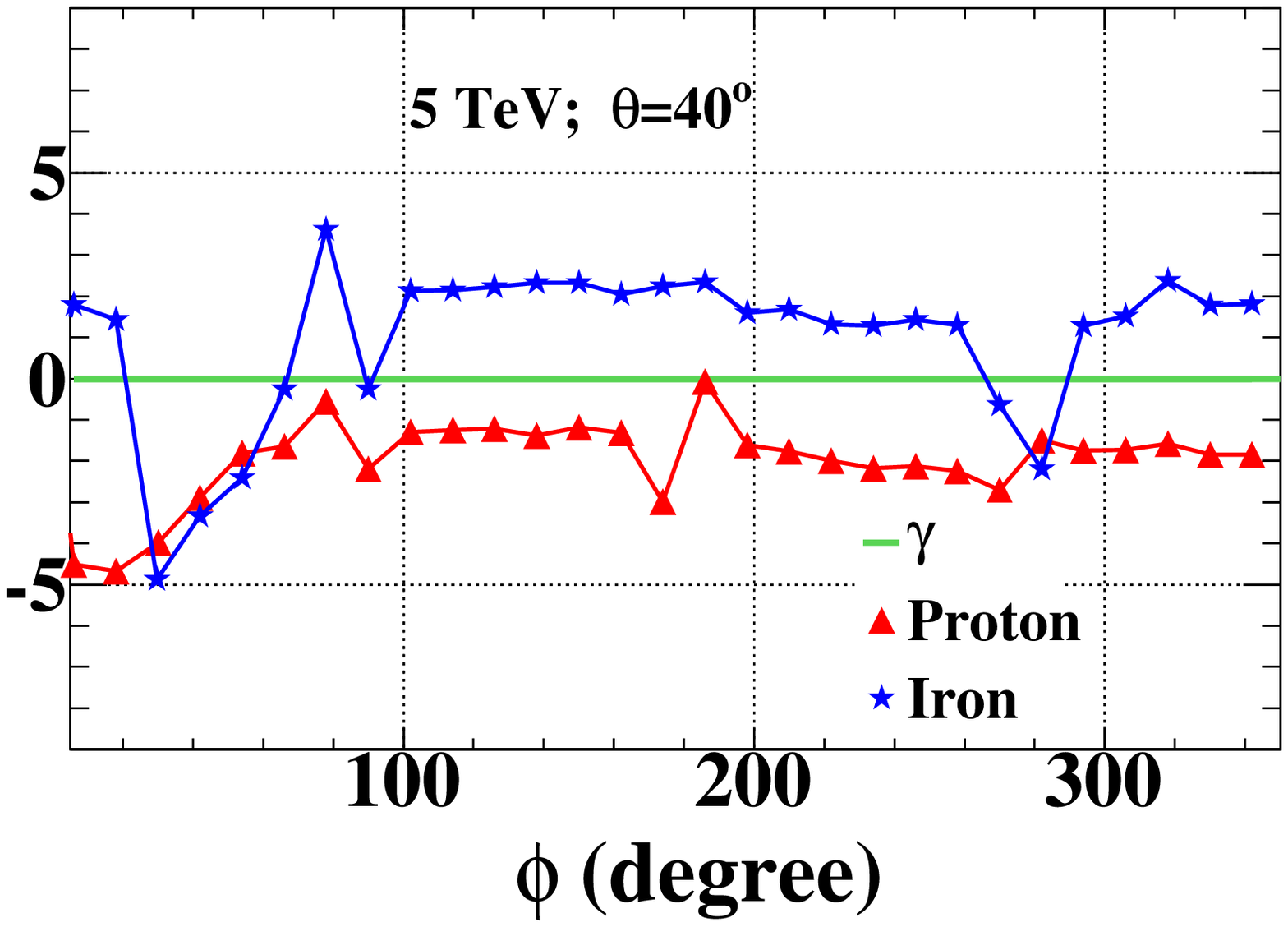}}
\caption{Percentage relative difference ($\Delta_N$) between the histograms 
of Cherenkov photons for $\gamma$-ray, proton and iron primaries.}
\label{fig5}
\end{figure}

To justify our explanation that two peaks of the azimuthal distribution of
Cherenkov photons in EAS are due to separation of electrons and positrons over 
the
azimuthal plane as a consequence of geomagnetic effect as mentioned earlier, 
in the Fig.\ref{fig3} we have shown the azimuthal distributions of electrons 
and positrons initiated by the 1 TeV $\gamma$-rays, as an example. This figure 
clearly shows that the 
left peak of azimuthal distribution of Cherenkov photons is due to electrons 
while right hand side peak is due to the contributions from positrons. The 
separation between electron peak and the positron peak is due to the 
geomagnetic effect or effect of the earth's magnetic field \cite{Homola}. The 
geomagnetic field deflects the electrons and positrons in opposite directions 
due to the opposite nature of their charge. Since the observational site in 
our study has a strong geomagnetic field (see the section \ref{sec2}), the 
separation is quite prominent. 

In the Fig.\ref{fig4} we have shown the comparison of the three primaries for 
a primary energy of 5 TeV at three different angles of inclination. From the 
figure it can be seen that the azimuthal distributions of Cherenkov photons due to the three primaries of same energy are not much different, but they are 
looked very much similar. However, to see the difference between these 
histograms clearly, we have calculated the percentage relative difference of 
the proton initiated and iron initiated histograms from that of $\gamma$-ray 
initiated histogram for the combination of energy and angle of incidence as
shown in the Fig.\ref{fig4}. The percentage relative difference has been 
obtained by using the following formula:
\begin{equation}
\Delta_{N} = \frac{N_{x} - N_{\gamma}}{N_{\gamma}}\times 100\%,
\label{eq1}
\end{equation}                 
where $\Delta_{N}$ is the percentage relative difference, $N_{x}$ and 
$N_{\gamma}$ are the number Cherenkov photons per shower at a given azimuthal
angle bin corresponding to the given primary and to the $\gamma$-ray primary 
respectively. From the  Fig.\ref{fig5} it can be seen that the percentage 
relative difference of the azimuthal distributions of Cherenkov photons 
initiated by the proton and iron primaries to that with initiated by 
the $\gamma$-ray 
is mostly limited to 5$\%$. More differences are basically noticeable for 
smaller azimuthal angles. However, with higher angle of incidence such 
noticeable differences persist for over all range of azimuthal angles.  

%However some differences can be observed below the 
%100$^{o}$azimuthal angle angle and in between 250$^{o}$ and 300$^{o}$.These regions where some amount of differences can be observed are basically the azimuthal angles corresponding to the two peaks of the histograms.Although no specific reasons can be attributed to these observed difference, but it is evident that the difference is prominent for those angular bins wherever the count of photon is either maximum or minimum. 

\subsection{Electron-positron asymmetry}
\begin{figure}[hbt]
\centerline
\centerline{
\includegraphics[scale = 0.29]{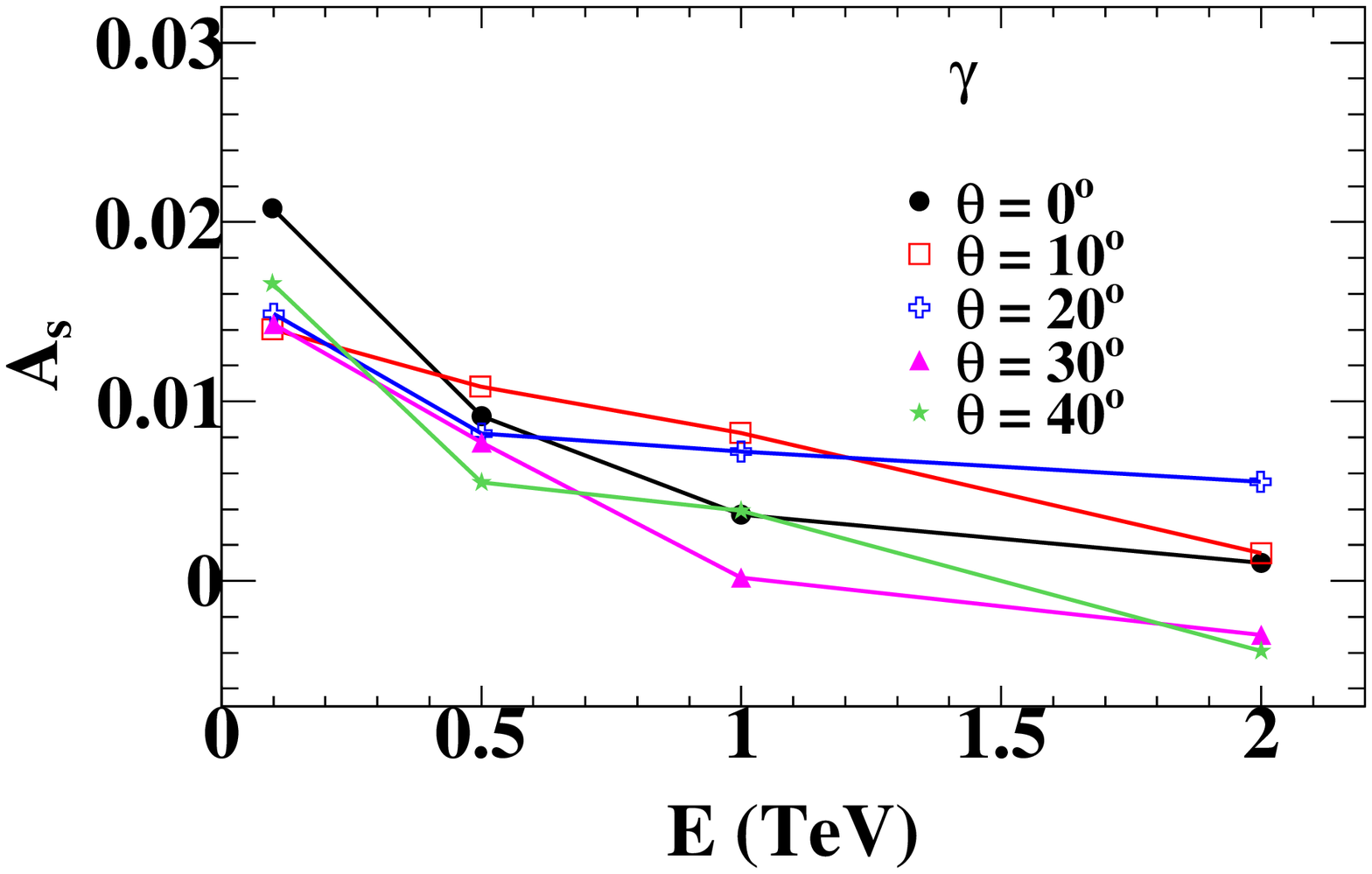}\hspace{-0.3cm}
\includegraphics[scale = 0.29]{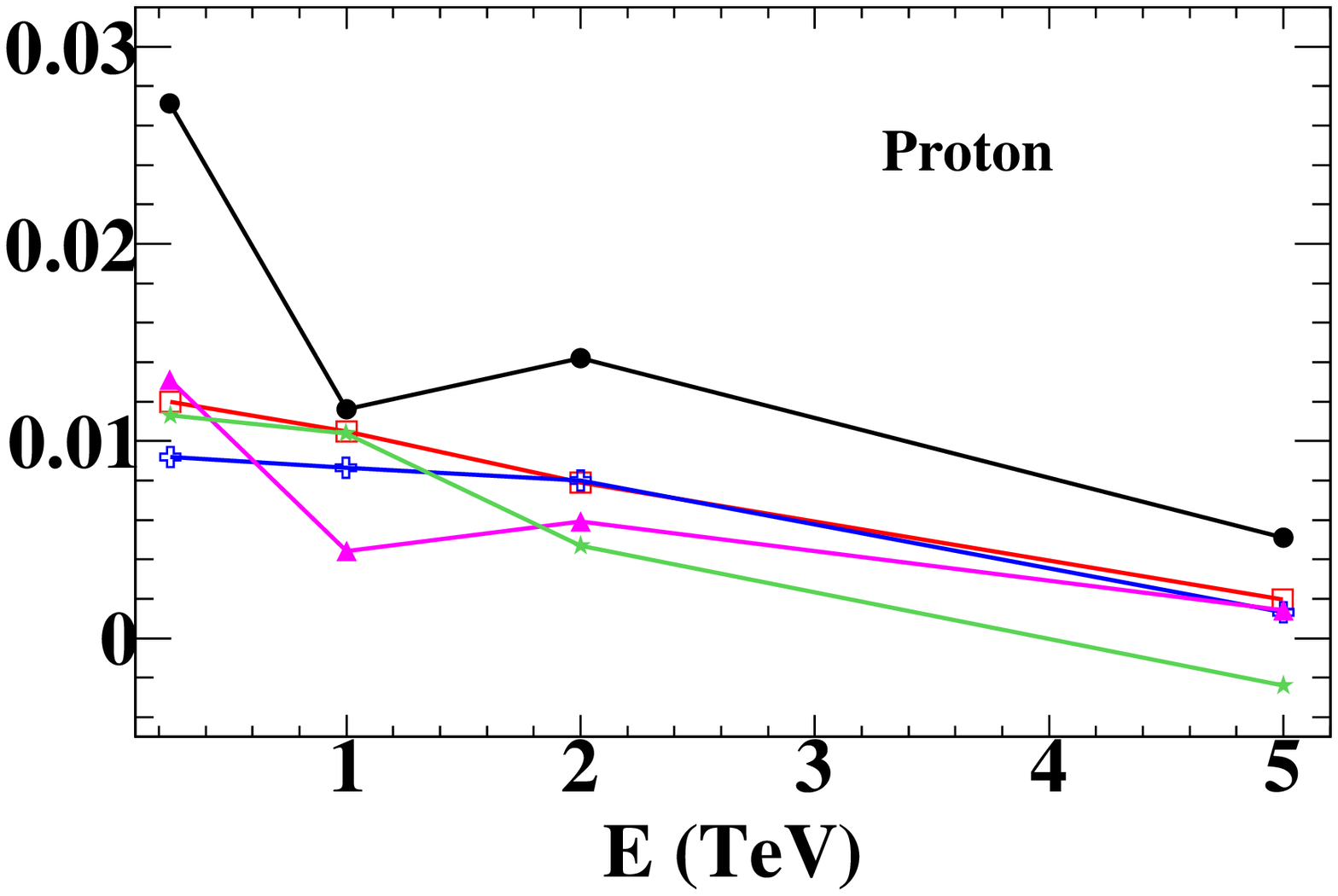}\hspace{-0.3cm}
\includegraphics[scale = 0.29]{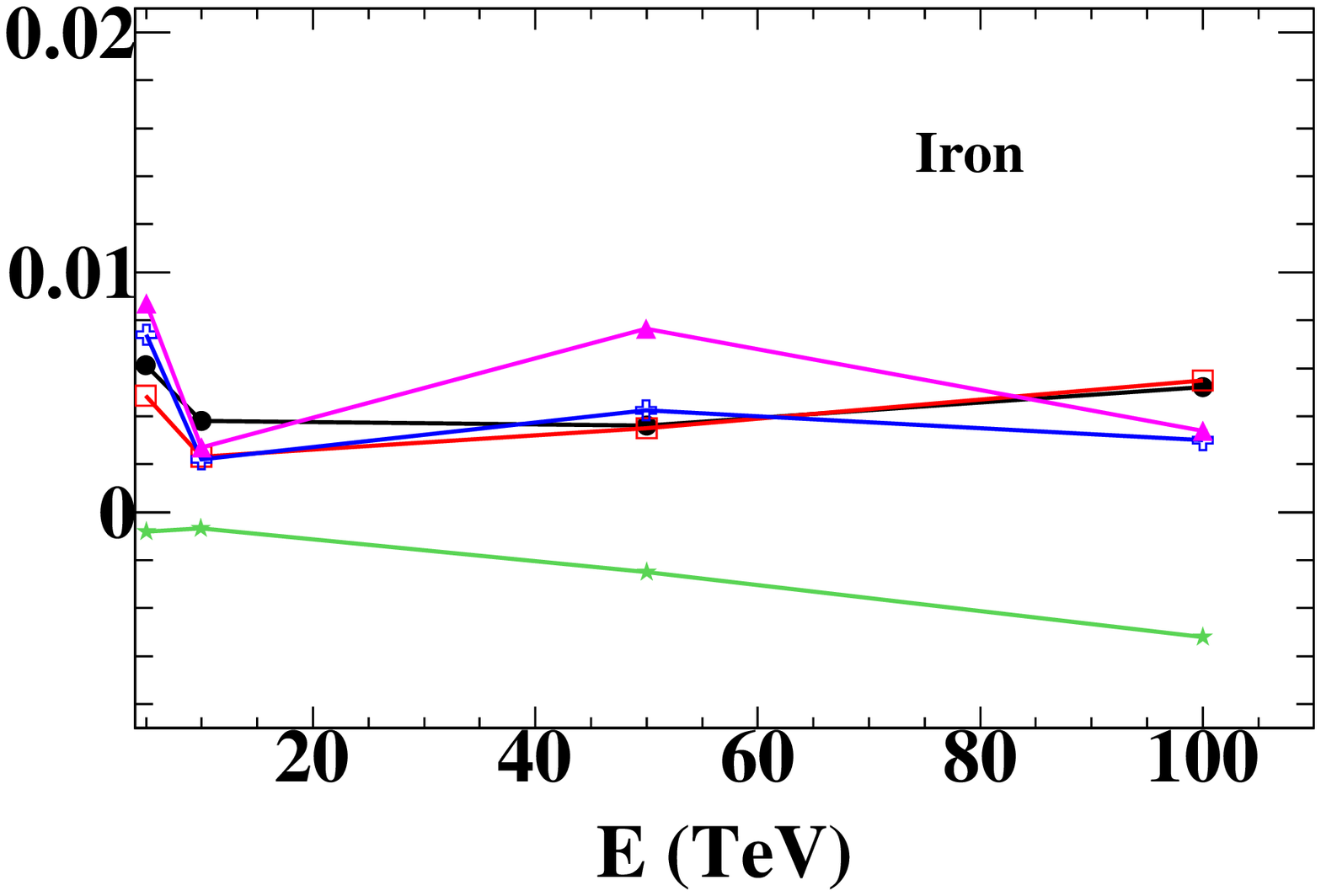}}
\caption{Electron-positron asymmetry as a function of energy for different 
values of angle of incidence initiated by $\gamma$-ray, proton and iron 
primaries.}
\label{fig6}
\end{figure}

\begin{figure}[hbt]
\centerline
\centerline{
\includegraphics[scale = 0.29]{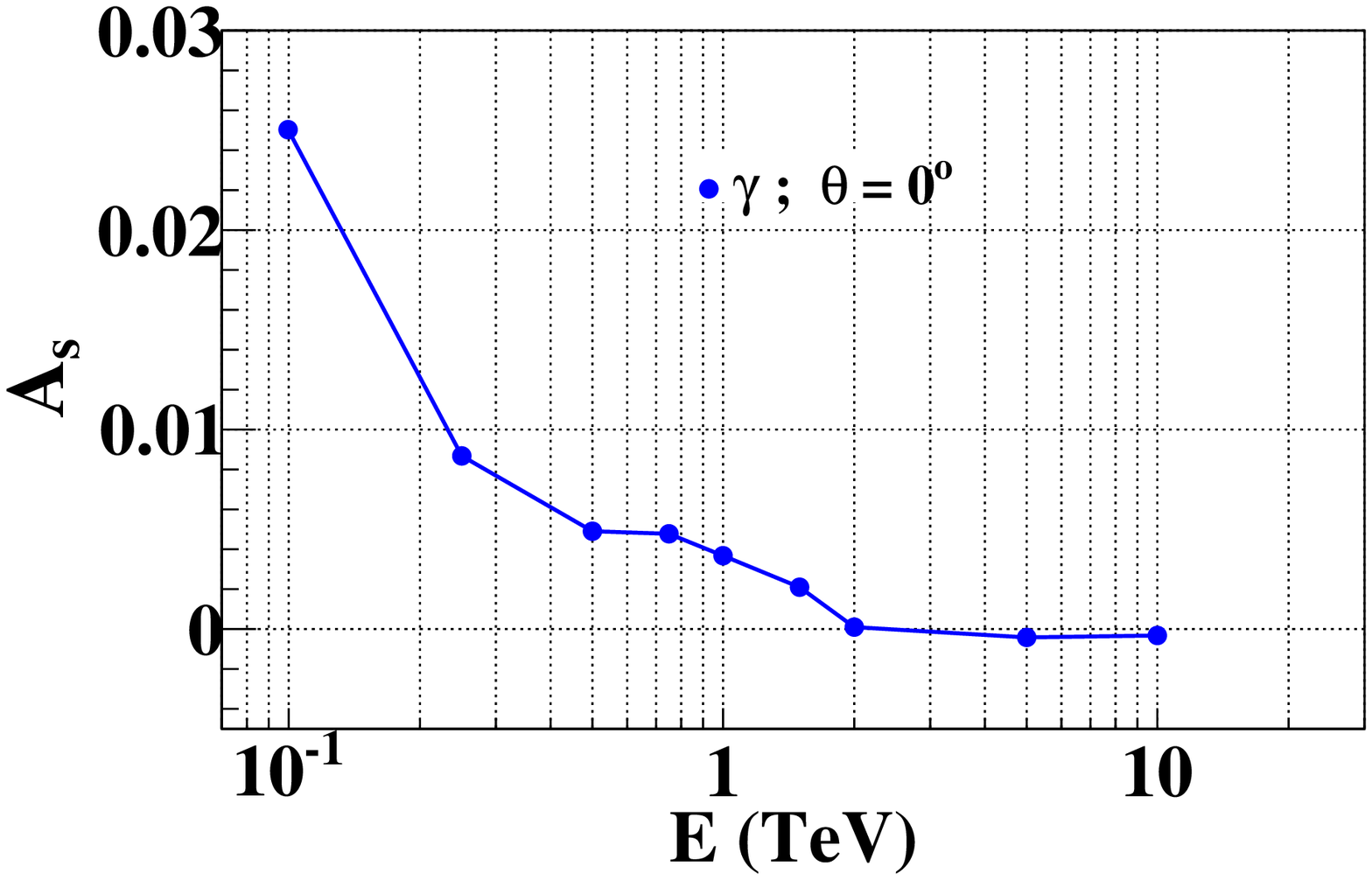}\hspace{-0.3cm}
\includegraphics[scale = 0.29]{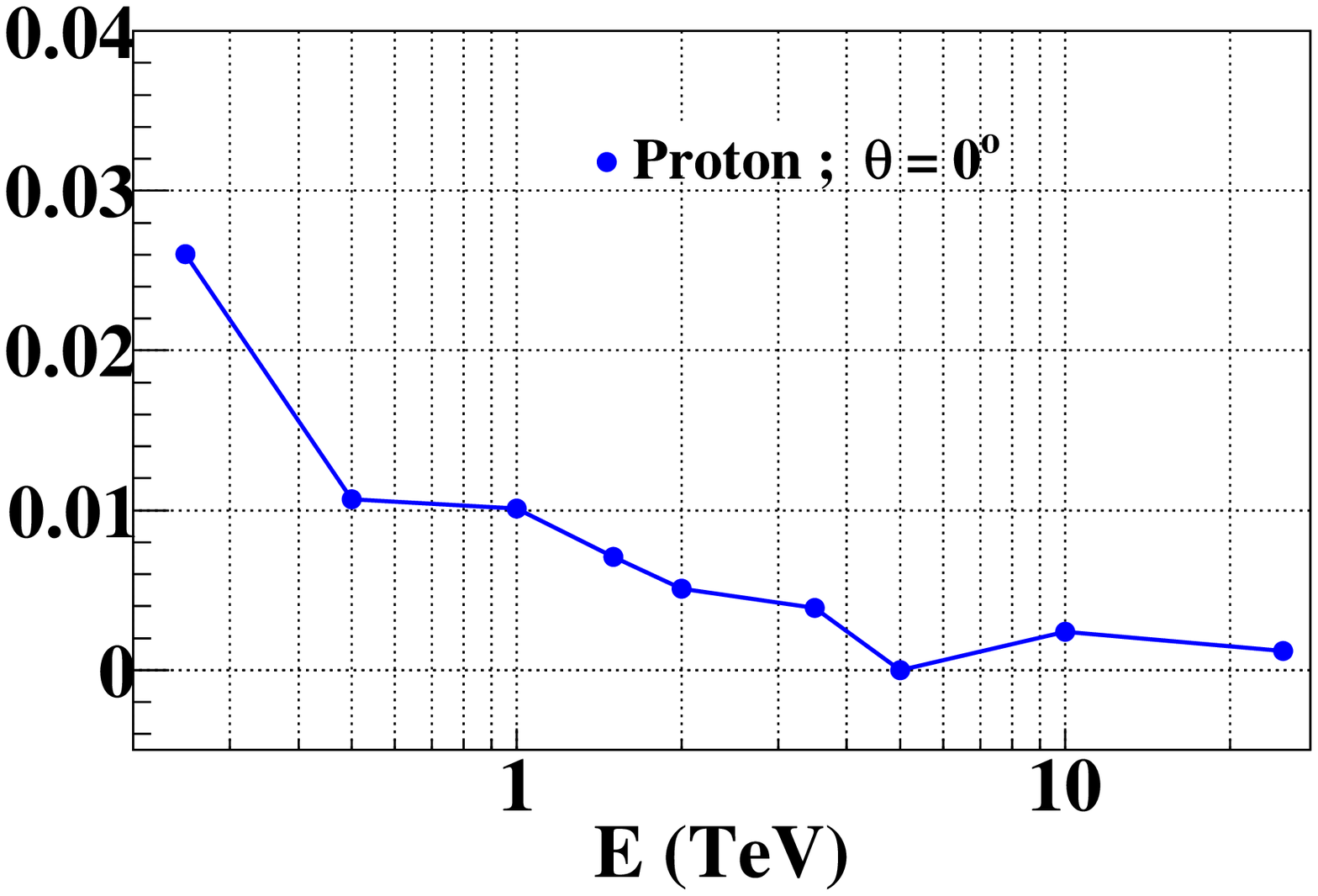}\hspace{-0.3cm}
\includegraphics[scale = 0.29]{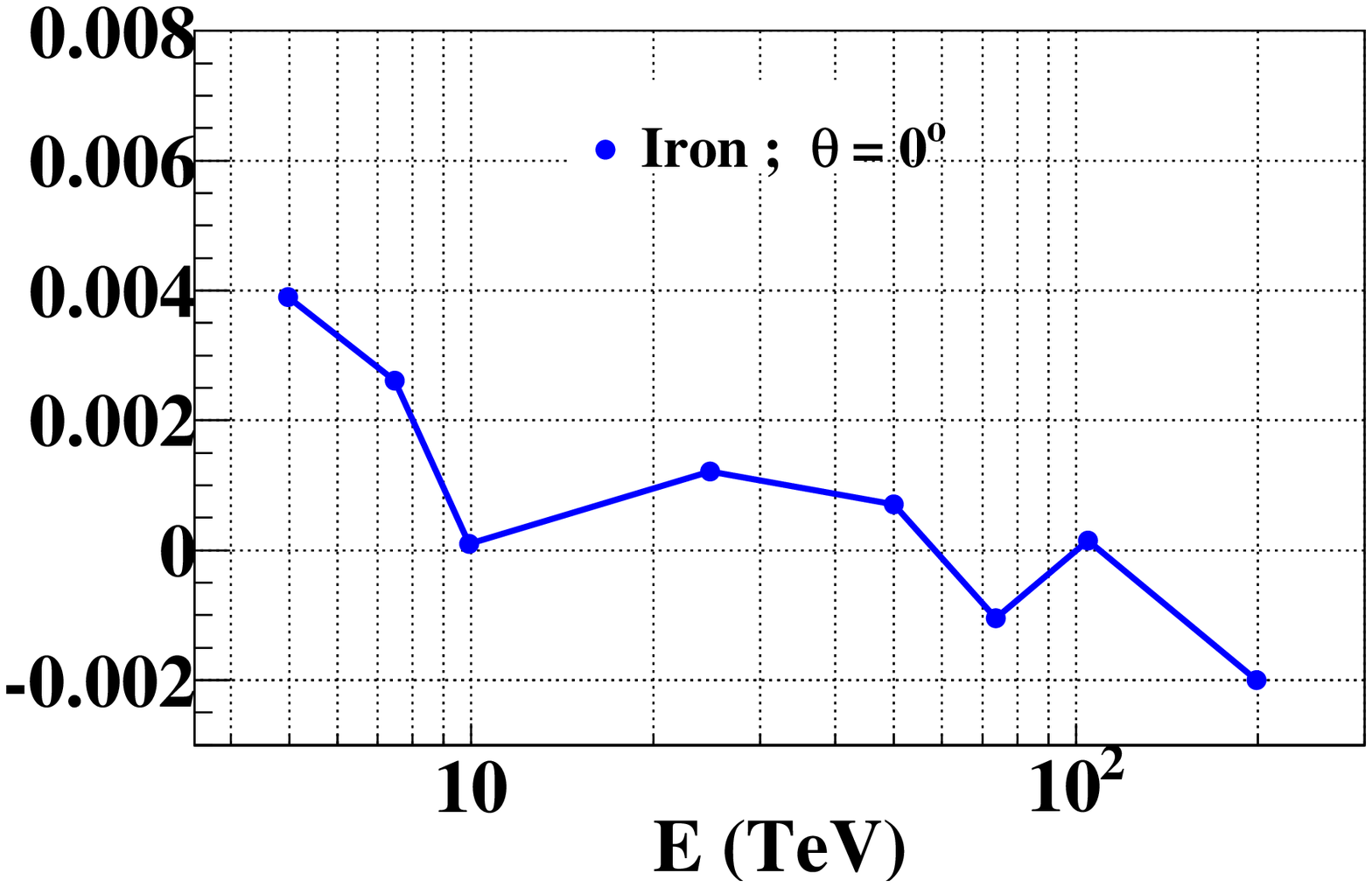}}
\caption{Electron positron asymmetry as a function of energy for vertically 
incident $\gamma$-ray, proton and iron primaries.}
\label{fig7}
\end{figure}

As already mentioned that due to the earth's magnetic field the Cherenkov 
photons produced by electrons and the Cherenkov photons produced by positrons 
are separated in the azimuthal plane which is manifested as two peaks in the 
azimuthal distribution profile of Cherenkov photons. In Fig.\ref{fig6} and 
Fig.\ref{fig7}, we have 
shown the electron-positron asymmetry as a function of energy for the 
$\gamma$-ray, proton and iron initiated showers. In all the cases we have seen 
positive asymmetry indicating the higher numbers of electrons in comparison to 
positrons. However, with the increase in energy, for all the 
primary particle and angle of incidence, we can see an increasing 
tendency towards symmetry. This is expected as higher energy particles are 
less affected by the earth's magnetic field. In the Fig.\ref{fig6}, we can see 
certain difference in the variation of asymmetry with energy for the iron 
primaries then in comparison to $\gamma$-ray and proton primaries. For iron primaries, 
the magnitude of asymmetry is less in comparison to $\gamma$-ray and proton
primaries. Also for the $\gamma$-ray and proton primaries we can observe a 
near linear decrease in the asymmetry whereas for iron it is not that smoothly 
varying (see Fig.\ref{fig7}). This is due to the complex nature of iron 
initiated showers than that initiated by other primaries.

\section{Summary and conclusion}
\label{sec4}
The objective behind this work was to study the distribution pattern of 
Cherenkov photons in the azimuthal plane to see whether the 
information about the azimuthal distribution can be used for $\gamma$-hadron 
separation along with the study of electron-positron asymmetry due to earth's 
magnetic field. The study has showed the double peak distribution of Cherenkov 
photons in the azimuthal plane. Also we have seen that the variations of the 
distribution is more sensitive to the parameter energy of the primary than the angle of incidence. This is expected, as with the energy of the incident 
particle the effect of earth's magnetic field also changes. Further, for the 
range of energy and angle of incidence combinations, that we have used in our 
study, azimuthal distributions of photons do not show any significant 
differences between the three primaries except at certain azimuthal angles. 
However, the variation of electron and positron asymmetry with energy shows 
certain differences in the three primaries- specially for the iron primaries.
This could be due to the simple fact that since iron is comparatively much 
heavier hadron primary, so its EAS is much complex than that of other 
primaries \cite{Hazarika, Das}. Hence, we can conclude that though the study 
reveals some interesting behaviours of Cherenkov photon distributions in the 
azimuthal plane, a more elaborate study covering a wider range of energy of 
the primaries and also inclusion of few more sensitive parameters may give a 
more insight in the $\gamma$-hadron separation possibility from such study. In 
future we plan to report the same.

\section*{Acknowledgments}
UGD is thankful to the Inter-University Centre for Astronomy and Astrophysics 
(IUCAA), Pune for hospitality during his visits as a Visiting Associate.

\end{document}